\documentclass[a4paper,11pt]{article}
\pdfoutput=1 

\usepackage{amsmath}
\usepackage{booktabs}
\usepackage{bm}
\usepackage{multicol}
\usepackage{multirow}
\usepackage{makecell}
\usepackage{caption}
\usepackage{subcaption}
\usepackage{graphicx}
\usepackage{xcolor}
\usepackage[utf8]{inputenc}
\newcommand{\UK}[1]{{\textcolor{black}{#1}}}

\DeclareUnicodeCharacter{2212}{-}

\usepackage{jcappub} 

\usepackage[T1]{fontenc} 

\title{Updated Constraints on Infrared Cutoff Models and Implications for Large-Scale CMB Anomalies}

\author{Ujjwal Upadhyay$^{1,2\,\ast}$,}
\emailAdd{$^\ast$ujjwalu@iisc.ac.in}
\affiliation{$^1$Department of Physics, Indian Institute of Science, Bangalore, India}
\author{Yashi Tiwari$^{2\,\dagger}$ and}
\emailAdd{$^\dagger$yashitiwari@alum.iisc.ac.in}
\affiliation{$^2$Astronomy and Astrophysics Group, Raman Research Institute, Bangalore, India}
\author{Tarun Souradeep$^{2\,\ddagger}$}
\emailAdd{$^\ddagger$tarun@rri.res.in}

\abstract{The nearly scale-invariant primordial power spectrum provides the standard initial conditions for cosmological perturbations. However, the largest scales remain only weakly constrained by CMB observations, leaving room for deviations such as an infrared (IR) cut-off. This possibility is further motivated by the persistence of large-scale CMB anomalies, most notably the low quadrupole power. In this work, we revisit several broad classes of phenomenologically motivated IR cut-off scenarios using parametrised functional forms of the primordial power spectrum. We confront these models with the latest CMB, BAO, and supernova data and derive updated constraints on the cut-off scale and associated features. Our results remain consistent with earlier studies, showing that although such models suppress power at low multipoles, the improvement in fit is marginal and does not overcome the associated parameter penalties. We therefore find no statistically significant evidence favouring IR cut-off models over the standard power-law spectrum with current data. We further explore the interplay between IR cut-off features and a possible increase in the reionisation optical depth, motivated by the recent CMB–BAO tension highlighted by DESI DR2 within the $\Lambda$CDM framework.  We show that the additional freedom introduced by large-scale suppression is generally insufficient to support a substantial increase in optical depth, owing to the weak statistical preference for suppressed large-scale temperature power. Finally, we examine the implications of IR cut-off models for large-scale CMB anomalies by analysing the corresponding anomaly statistics within a Bayesian framework.}

\begin{document}
\maketitle
\flushbottom

\section{Introduction}
\label{sec:intro}
The epoch of cosmological inflation, marked a period of rapid, nearly exponential expansion in the early Universe, is a central component of the standard Lambda--Cold Dark Matter ($\Lambda$CDM) cosmological model. Originally proposed to address several fundamental problems of the hot Big Bang framework, inflation has been remarkably successful, not only in resolving these conceptual issues but also in providing a natural mechanism for generating primordial curvature perturbations \cite{PhysRevD.23.347, 1982PhLB..108..389L}. These primordial fluctuations subsequently evolve into the observed anisotropies in the Cosmic Microwave Background (CMB) and the large-scale structure of the Universe \cite{Mukhanov:1981xt,Tegmark:2004qd}. The initial conditions for cosmological perturbations generated during inflation are characterised by the primordial power spectrum (PPS) of curvature perturbations. A generic prediction of the simplest inflationary models is a nearly scale-invariant scalar power spectrum, commonly parametrised by a power law,
$P_s(k) = A_s (k/k_*)^{n_s - 1}$,
where $A_s$ and $n_s$ denote the amplitude and spectral index, respectively, and $k_*$ is a chosen pivot scale. These two quantities constitute part of the six free parameters of the standard $\Lambda$CDM model. This simple form of the primordial power spectrum is in excellent agreement with observations of the CMB temperature and polarisation power spectra, which tightly constrain both $A_s$ and $n_s$ within the $\Lambda$CDM framework \cite{Planck:2018vyg, Planck:2018jri}. Nevertheless, in comparison to other cosmological parameters, these inflationary parameters are more phenomenological in nature: they specify the initial conditions of the perturbations and acquire physical significance only through their imprint on observable quantities as the Universe evolves within a given cosmological model.
These considerations motivate going beyond the simplest power-law description to allow for scale-dependent features in the primordial power spectrum. One of the simplest extensions is to promote the spectral index to a scale-dependent quantity, $n_s \rightarrow n_s(k)$, which can be parametrised in terms of its running and higher-order derivatives. While this approach is more flexible than a pure power law, capturing non-trivial features typically requires introducing several additional parameters. As a result, even when such extensions yield modest improvements in the quality of fit, they may be statistically disfavoured once the increased model complexity is taken into account.
\par A complementary and more flexible approach to extracting information about the primordial power spectrum relies on reconstruction techniques, which infer PPS directly from observations in a largely model-agnostic manner \cite{Tegmark:2002cy, Bridle:2003sa, Leach:2005av, Tocchini-Valentini:2005mzh, Hazra:2013nca, Planck:2018jri, Handley:2019fll}. Such analyses consistently indicate that CMB observations are sensitive only to a limited range of scales, leaving the primordial power on the largest scales weakly constrained. This naturally motivates the exploration of large-scale primordial features that represent deviations from a simple power-law form in the infrared regime. Indeed, several model-independent reconstructions of the primordial power spectrum have reported a preference for suppressed power on the largest scales, consistent with the presence of an effective infrared cutoff. For example, analyses based on the modified Richardson–Lucy deconvolution algorithm have shown that the best-fit reconstructed primordial power spectrum from CMB data exhibits an infrared cutoff, followed by a localized enhancement and decaying oscillatory features \cite{Shafieloo:2003gf, Shafieloo:2007tk, Hazra:2014jwa, Sohn:2022jsm}. Nevertheless, the statistical significance of these features remains limited, largely due to the strong impact of cosmic variance on the largest angular scales probed by the CMB.
\par From a phenomenological standpoint, the microphysics of inflation, including the mechanism responsible for its onset and the state of the Universe prior to inflation, remains weakly constrained, largely because CMB observations have limited sensitivity to the largest scales \cite{Knox:1995dq, Planck:2018jri, Planck:2019evm}. While current data are well described by single-field slow-roll inflation \cite{Martin:2013tda}, a wide range of scenarios involving non-standard pre-inflationary dynamics or departures from simple slow-roll dynamics on poorly constrained scales remains viable. Among these, particular attention has been given to models that generate large-scale features in the primordial power spectrum. Such features may include an infrared cut-off or a localised suppression of power on the largest scales, arising from mechanisms such as a pre-inflationary radiation-dominated phase, an initial fast-roll stage, a step or inflection point in the inflaton potential, or more general multi-field dynamics  \cite{Starobinsky:1992ts, Contaldi:2003zv, Cline:2003ve, Kawasaki:2003dd, Powell:2006yg, Jain:2008dw, Jain:2007au, Qureshi:2016pjy, Ashoorioon:2014yua, Handley:2014bqa, Hazra:2010ve, Braglia:2021sun, Tiwari:2022zzz, Braglia:2020fms}. Such constructions are largely motivated by observed anomalies in the CMB temperature anisotropy power spectrum, most notably the anomalously low quadrupole, first reported by COBE and subsequently confirmed by WMAP and Planck \cite{Bennett:1996ce, WMAP:2008ttx, Schwarz:2015cma, Planck:2019evm}. Although the low quadrupole anomaly does not carry strong statistical significance on its own, it has nevertheless motivated investigations of phenomenological models—such as IR cutoff scenarios—that can yield an improved fit to the large-scale CMB data. In practice, most studies find that a suppression of power on the largest scales leads to only modest improvements in the goodness of fit. Moreover, such scenarios are typically penalised in model selection analyses due to the introduction of additional parameters. Any gain from improved agreement with the observed large-scale temperature anisotropies is further diminished by the large cosmic variance inherent at low multipoles, which limits the statistical weight of these scales.
\par In this work, we revisit a set of IR cutoff scenarios, parametrised through different functional forms of the primordial power spectrum, corresponding to a range of phenomenologically motivated inflationary models. Our goal is to reassess the current viability and relative performance of these scenarios compared to the standard power-law $\Lambda$CDM model, in light of recent cosmological observations, including measurements of the CMB, Baryon Acoustic Oscillations (BAO), and Type Ia supernovae (SNIa). To this end, we adopt a Bayesian framework that enables both parameter inference and model comparison, employing robust statistical analyses to constrain the standard cosmological parameters as well as the additional IR cutoff parameters. We use the latest high-precision CMB data from Planck, the Atacama Cosmology Telescope (ACT), and the South Pole Telescope (SPT). While ACT and SPT primarily probe smaller angular scales (higher multipoles), well away from the infrared cutoff scale, their inclusion is particularly relevant for IR cutoff models that predict oscillatory features extending to smaller scales beyond the cutoff. Henceforth, we also explore whether such features can leave observable imprints in the multipole range accessed by ACT and SPT, thereby providing complementary constraining power to the large-scale CMB measurements.
\par Another motivation for revisiting IR cutoff scenarios comes from the recently reported discrepancy between CMB and BAO observations following the DESI release within the $\Lambda$CDM framework \cite{DESI:2025zgx}. This tension has been interpreted in several studies as hinting at departures from the standard cosmological model, including the possibility of a dynamical dark energy component \cite{DESI:2025zgx, DESI:2025fii}. It has also been suggested that the inconsistency can be alleviated by allowing for a higher value of the reionisation optical depth, $\tau_{\rm reio}$, which is otherwise tightly constrained by large-scale CMB polarisation measurements \cite{Jhaveri:2025neg, Sailer:2025lxj}. In this context, IR cutoff models of the primordial power spectrum have attracted renewed attention because of their potential degeneracies with reionisation signatures in the CMB. In particular, it has been pointed out that large-scale suppression in the primordial power spectrum could bias the inference of $\tau_{\rm reio}$, permitting larger values when confronted with cosmological data. Motivated by these considerations, we perform a comprehensive analysis to assess whether IR cutoff features can indeed play a meaningful role in reconciling the CMB--BAO discrepancy within the $\Lambda$CDM framework. Our analysis, however, demonstrates that IR cutoff models cannot rescue the standard $\Lambda$CDM scenario in this context. The suppression features are tightly constrained by large-scale CMB temperature anisotropy data, leaving little room for such models to bias the inference of reionisation optical depth.
\par In the final part of this work, we examine the implications of IR cutoff models for large-scale CMB anomalies, with particular emphasis on the low quadrupole power and the parity asymmetry observed at low multipoles and explore possible correlations between them. We adopt a Bayesian perspective to analyse these anomalies by studying the predictions of both IR cutoff models and the standard power-law scenario for suitably defined anomaly statistics. These model-dependent predictions are then compared with those of a `true theory' given the observations, which is obtained in a model-agnostic manner under the minimal assumptions of statistical isotropy and Gaussian random initial conditions. 
This approach provides a complementary Bayesian framework for assessing the significance of large-scale CMB anomalies using anomaly statistics, \UK{that} are traditionally discussed within a frequentist context.
\par This paper is organised as follows. In Sec.~\ref{sec:second}, we describe the functional forms of the primordial power spectra for the IR cutoff models considered in this work. In Sec.~\ref{sec:third}, we outline the methodology and the data sets employed in the analysis. The results and their implications are presented in Sec.~\ref{sec:fourth}. Finally, we summarise our findings and discuss their broader implications in Sec.~\ref{sec:five}.

\section{Infrared Cut-off Models}
\label{sec:second}
\begin{figure}[h]
    \centering
     \includegraphics[scale=0.9]{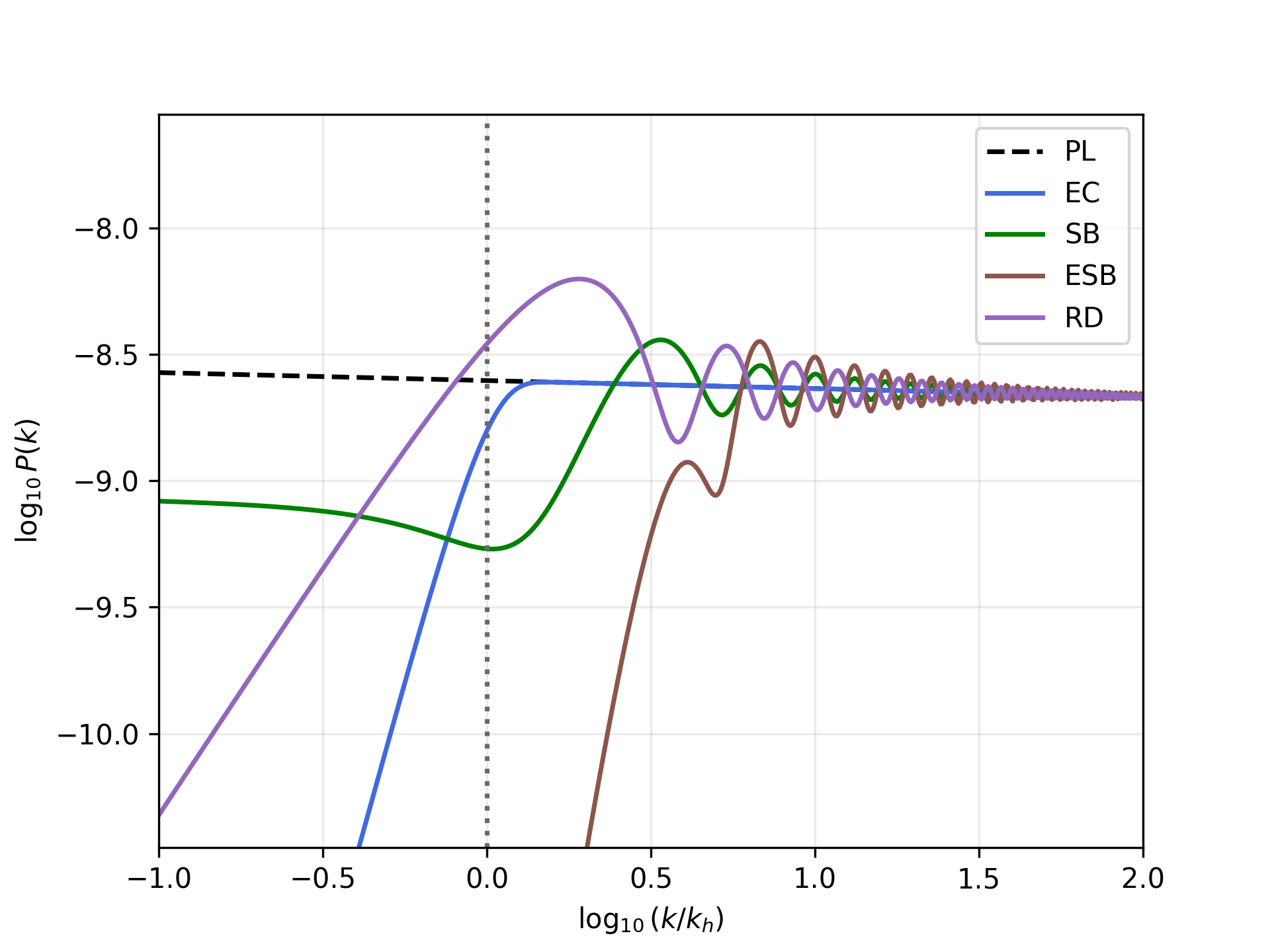}
     \caption{Primordial power spectra for the different IR cutoff scenarios discussed in Sec.~\ref{sec:second}, shown alongside the reference power-law spectrum (black dashed) for comparison. The vertical line marks the scale $k_h = 2.25 \times 10^{-4}\,\mathrm{Mpc}^{-1}$, corresponding to the horizon scale of the best-fit $\Lambda$CDM model.
 }
     \label{fig: IR-models-pps}
\end{figure}

The primordial power spectrum as the initial condition for the growth of perturbations assumes more information than what we can test from the current observations. For example, it assumes that the curvature perturbations are Gaussian, although current observations place only weak limits on possible non-Gaussian deviations.  Further, the standard initial condition assumes a power law with a constant red tilt on all scales, while the CMB observations constrain only a small window of the cosmological scales. The IR cutoff models with suppressed power on large scales are especially interesting in the context of large-scale CMB anomalies. In this section, we describe the form of some IR cutoff models of the primordial power spectrum of scalar curvature that arise in various physical models of inflation. We consider the following primordial power spectra as discussed in \cite{Sinha:2005mn}:

\begin{itemize}
    \item {\bf Power Law Model (PL):} Power law is the standard nearly scale-invariant initial condition predicted by the single-field inflationary models. 
    
    \begin{equation}
        \mathcal{P}_{\mathcal{R}}(k) =\mathcal{P}_0(k) = A_s \left( \frac{k}{k_*} \right)^{n_s - 1}
    \end{equation}
    Here, $A_s$ and $n_s$ are the amplitude and tilt of the spectrum, respectively, and $k_*=0.05\;\rm Mpc^{-1}$ is the pivot scale with the value set by the COBE normalization convention. This model has two free parameters -- $A_s$ and $n_s$, which are two of the six parameters of the $\Lambda$CDM model.

    \item {\bf Exponential Cutoff Model (EC):} This is a simple extension of the power law form with two additional parameters compared to the $\Lambda$CDM model given by the functional form \cite{Contaldi:2003zv, Cline:2003ve}
    \begin{equation}
       \mathcal{P}_{\mathcal{R}}(k) = \mathcal{P}_0(k) \left[ 1 - e^{-(k/k_c)^{\alpha}} \right]
    \end{equation}
    Here, $k_c$ sets the cutoff scale and $\alpha$ controls the sharpness of the cutoff. On very small scales ($k>>k_c$), it approaches the standard power law, while on large scales ($k<<k_c$), the power is suppressed to zero.

    \item {\bf Starobinsky Break Model (SB):} This model predicts a smooth large-scale suppression of power followed by damped oscillatory features at smaller scales. \UK{Such behaviour arises generically in inflationary scenarios where the inflaton potential remains continuous but exhibits a sharp change in its first derivative, leading to a brief departure from slow-roll evolution \cite{Starobinsky:1992ts}.} The functional form of the primordial power spectrum in the SB model is given by, 
    \begin{equation}
        \mathcal{P}_{\mathcal{R}}(k) = P_0(k)\, \mathcal{D}^{2}(y, R_*)
    \end{equation}
    where
    \begin{align}
    \mathcal{D}^2(y, R_*) =\;& 1 - 3 (R_* - 1)\,\frac{1}{y}
    \left( \left(1 - \frac{1}{y^2}\right)\sin(2y) + \frac{2}{y}\cos(2y) \right) \nonumber \\
    & + \frac{9}{2}(R_* - 1)^2 \frac{1}{y^2}\left(1 + \frac{1}{y^2}\right)
    \left( 1 + \frac{1}{y^2} + \left(1 - \frac{1}{y^2}\right)\cos(2y) - \frac{2}{y}\sin(2y) \right)
    \end{align}
    Here, $P_0(k)$ is the power law model.

    \item \textbf{Exponential--Starobinsky Model (ESB):}
This model combines the features of an exponential cutoff with a Starobinsky-type break, resulting in a suppression of power at the largest scales followed by a localized enhancement (bump) at intermediate scales \cite{Sinha:2005mn}. The primordial curvature power spectrum is given by
\begin{equation}
\mathcal{P}_{\mathcal{R}}(k) = P_0(k)\left[1 - \exp\!\left(-\left(\frac{\epsilon k}{k_c}\right)^{\alpha}\right)\right]\,\mathcal{D}^2(y, R_*),
\end{equation}
where the exponential factor introduces a smooth cutoff at wavenumbers $k \lesssim k_c$, while the transfer function $\mathcal{D}(y, R_*)$ encodes the Starobinsky break and generates oscillatory features around the cutoff scale. The additional parameter $\epsilon$ controls the relative position and sharpness of the cutoff with respect to the Starobinsky feature, allowing greater flexibility in tuning the interplay between suppression and bump in the primordial spectrum. However, in this work, we use $\epsilon=0.2$ throughout the analysis, following the discussions of \cite{Sinha:2005mn}.

    \item \textbf{Pre-inflationary Radiation Domination Model (RD):}
    This class of phenomenological models assumes a phase of radiation domination preceding the onset of inflation, which modifies the initial conditions for cosmological perturbations relative to the standard Bunch--Davies vacuum. As a consequence, the primordial modes that exit the horizon during the early stages of inflation retain memory of this pre-inflationary phase, leading to characteristic large-scale features in the primordial power spectrum \cite{PhysRevD.26.1231, Powell:2006yg}.

    The resulting curvature power spectrum can be written as
    \begin{equation}
    \mathcal{P}_{\mathcal{R}}(k)= \mathcal{P}_0(k) \,
    \frac{1}{4 y^4} \,
    \left| e^{-2 i y} (1 + 2 i y) - 1 - 2 y^2 \right|^2 ,
    \label{eq:RD}
    \end{equation}
    where $y \equiv k/k_c$, and $k_c$ denotes the characteristic cutoff scale associated with the transition between the radiation-dominated phase and inflation. This power spectrum exhibits a suppression of power at the largest scales, followed by a localized enhancement and oscillatory features at intermediate wavenumbers. \UK{We also note that analogous large-scale dynamics can arise in alternative phenomenological scenarios featuring a fast-roll (kinetic-dominated) phase preceding the onset of slow-roll inflation \cite{Contaldi:2003zv, Destri:2009hn}. Since such models effectively lead to similar large-scale features, we do not treat kinetic-dominated scenarios as a separate case in this analysis.}
\end{itemize}
Figure~\ref{fig: IR-models-pps} shows the primordial power spectra corresponding to the different models discussed above, illustrating their distinct characteristic features, all of which include a suppression of power on the largest scales.

\section{Data and Methodology} 
\label{sec:third}
In this section, we outline the cosmological datasets and the methodology used in our analysis. To constrain the IR cutoff models, we employ the latest observations from the CMB, BAO, and supernovae. Our approach is based on Bayesian parameter estimation using Markov Chain Monte Carlo (MCMC) sampling, along with model comparison. Furthermore, we derive the posterior distributions of key summary statistics associated with large-scale CMB anomalies—particularly those linked to the low quadrupole anomaly—to provide a Bayesian perspective on their statistical significance.        
\subsection{Datasets}
\label{sec: data}
The primary impact of IR cut-off features in the primordial power spectrum manifests at low multipoles in the angular CMB power spectrum. However, most phenomenological IR cut-off models are generically accompanied by oscillatory features on smaller scales, arising as a consequence of the sharp suppression at large scales, as illustrated in Fig.~\ref{fig: IR-models-pps}. These small-scale oscillations are particularly relevant in light of recent ACT and SPT measurements, that extend CMB anisotropy observations to significantly higher multipoles and could therefore be sensitive to such signatures \cite{ACT:2025fju, SPT-3G:2025bzu}. As a result, the dominant constraints on these models are provided by the CMB data. Once confronted with observations, however, the model parameters become correlated, such that constraints on one aspect of the model propagate to the others. For this reason, in addition to the CMB data, we include complementary information from BAO and SNIa observations in order to obtain robust and tightened constraints on the model parameter space. We use the following datasets and their combinations in our analysis:
\begin{itemize}
    \item {\bf Planck:} We use the Planck low-$\ell$ {\tt TT} and {\tt EE} (sroll2) likelihoods \cite{Planck:2018nkj, Pagano:2019tci, Delouis:2019bub}. At high multipoles, we adopt the Planck Public Release 3 (PR3) Plik likelihood, restricting to $\ell<1000$ for {\tt TT} and $\ell<600$ for {\tt TE} and {\tt EE}, following \cite{ACT:2025fju}, referred to as {\tt Planck$_\text{cut}$} in our analysis. This choice allows for a consistent combination with recent ACT measurements, which are incorporated at higher multipoles where ACT provides comparable or stronger constraining power than Planck. We further include CMB lensing data from Planck PR4, which is more constraining than the PR3 lensing likelihood \cite{Carron:2022eyg}. We will refer to it as {\tt PR4 lensing}.
    \item {\bf ACT DR6:} We use the latest ACT DR6 likelihood, which includes the {\tt TT}, {\tt TE}, and {\tt EE} power spectra \cite{ACT:2025fju}, in combination with the {\tt Planck$_\text{cut}$} and PR4 lensing data. We refer to this likelihood as {\tt ACT}.
    \item {\bf SPT-3G D1:} We use the latest SPT-3G D1 likelihood, which includes the {\tt TT}, {\tt TE}, and {\tt EE} power spectra \cite{SPT-3G:2025bzu}, hereafter referred to as {\tt SPT}. This dataset is combined with {\tt PlanckACT$_\text{cut}$} and {\tt ACT}. Since the SPT-3G and ACT DR6 measurements are uncorrelated, their likelihoods can be directly combined. Accordingly, our analysis employs the joint dataset {\tt PlanckACT$_\text{cut}$ $+$ ACT $+$ SPT $+$ PR4 lensing}, incorporating the most recent and complementary CMB observations across a wide range of angular scales.
    \item {\bf DESI DR2:} This is the DESI DR2 measurements of the BAO feature in the galaxy distribution. It consists of angular scales of transverse BAO and the radial BAO feature measured at 6 effective redshifts, and angle-averaged isotropic BAO at one effective redshift \cite{DESI:2025zgx}. The uncertainty in the measurement is provided in the form of a covariance matrix, and the likelihood is implemented in the \texttt{Cobaya}, a statistical code for Bayesian analysis \cite{Torrado:2020dgo}. 
    \item {\bf PantheonPlus:} It consists of redshifts and magnitude measurements of 1701 Type Ia supernovae in the redshift range,  $z \in [0.01, 2.26]$ \cite{Scolnic:2021amr}. The data provide both uncalibrated and SH0ES-calibrated magnitudes; however, the {\tt PantheonPlus} likelihood implemented in {\tt Cobaya} uses uncalibrated supernova magnitudes.
\end{itemize}
Since we are also interested in investigating how individual datasets affect the constraints on IR cutoff models, we consider various dataset combinations as described below:
\begin{table}[h]
\centering
\renewcommand{\arraystretch}{1.9}
    \centering

    \begin{tabular}{cp{0.1cm}c}
    \hline \hline
        \textbf{Combination} && \textbf{Likelihoods} \\
    \hline 
         CMB&& {\tt low-$\ell$ TT \& EE$+$ PlanckACT$_\text{cut}$ $+$ ACT $+$ SPT $+$ PR4 lensing } \\
CMB+BAO+SN&& { CMB \tt $+$ DESI DR2 $+$ PantheonPlus} \\
CMB (no low-$\ell$ EE) && {CMB \tt $-$ low-$\ell$ EE} \\
\hline 
    \end{tabular}
    \caption{Various combinations of the datasets used in the analysis.}
    \label{tab:datasets}
\end{table}\\

\begin{table}[tbh]
\centering
\renewcommand{\arraystretch}{1.8}
\setlength{\tabcolsep}{10pt} 
\begin{tabular}{p{7cm} l c p{2.5cm}} 
\hline \hline
\textbf{Description} & \textbf{Parameter} & & \textbf{Prior} \\
\hline
Physical baryon density & $\Omega_b h^2$  & & $\mathcal{U}(0.005,\,0.1)$ \\
Physical cold dark matter density & $\Omega_c h^2$  & & $\mathcal{U}(0.001,\,0.99)$ \\
Optical depth to reionization & $\tau_{reio}$  & & $\mathcal{U}(0.01,\,0.8)$ \\
Angular scale of sound horizon& $100\theta_s$  & & $\mathcal{U}(0.5,\,10)$ \\
Amplitude of primordial power spectrum & $\ln (10^{10}A_s)$  & & $\mathcal{U}(1.6,\,3.9)$ \\
Scalar spectral index & $n_s$  & & $\mathcal{U}(0.8,\,1.2)$ \\
\hline
Infrared cutoff scale  & $\ln k_c$  & & $\mathcal{U}(-12.0,\,-7.0)$ \\
Sharpness of the cutoff & $\alpha$  & & $\mathcal{U}(0.1,\,10.0)$ \\
Starobinsky parameter & $R_*$  & & $\mathcal{U}(0.01,\,1.0)$ \\
\hline
\end{tabular}
\caption{Priors adopted for the standard cosmological parameters and the IR cutoff model parameters in the MCMC analysis.}
\label{tab:priors}
\end{table}

\subsection{Methodology}
As outlined in the previous section, this work is structured around three main objectives: (i) assessing the current status of IR cutoff models using recent observational datasets, (ii) investigating the implications of IR cutoff features for the CMB–BAO discrepancy, and (iii) examining their consequences for large-scale CMB anomalies. All analyses are carried out within a Bayesian framework, involving posterior distribution sampling for parameter estimation as well as statistical model comparison. The methodology adopted for the first two objectives follows the standard approach and is described in detail in Sec.~\ref{sec:method-1}. The study of the impact of IR cutoff models on large-scale CMB anomalies, however, employs a complementary analysis strategy tailored to anomaly-related statistics, and the corresponding methodology is presented in Sec.~\ref{sec:method-2}.

\subsubsection{MCMC Analysis and Model Comparison }
\label{sec:method-1}
We carry out a Bayesian analysis of the IR cutoff models introduced in Sec.~\ref{sec:second} using MCMC sampling. The CMB power spectra are generated from the primordial power spectrum using the Boltzmann solver {\tt CLASS} \cite{Diego_Blas_2011}. Parameter inference is carried out with {\tt Cobaya}, employing the Metropolis–Hastings algorithm for sampling the posterior distribution \cite{Torrado:2020dgo}.
For the priors, we use uniform distribution on the six $\Lambda$CDM parameters as well as on the additional parameters governing the strength and location of the IR cutoff, as summarized in the Table \ref{tab:priors}. It should be noted that the parameters describing the primordial power spectra,\{$A_s, n_s, k_c, \alpha, R_*$\}, do not exactly have the same meaning due to the difference in the functional form of the PPS. For the parameters $A_s$ and $k_c$, we have used a log-uniform prior.
For each model, we run 16 parallel chains and stop when the desired convergence is achieved. Convergence is tested with the standard Gelman–Rubin criterion, requiring $R-1 < 0.02$ \cite{Gelman:1992zz}. To evaluate the performance of the IR cutoff models, we compare them against the baseline power-law primordial spectrum. We compute the best-fit $\chi^2$ values to quantify the goodness of fit and evaluate the relative improvement with respect to the power-law case. In addition, we employ the \UK{Akaike Information Criterion (AIC) \cite{BurnhamAnderson2002}},
\begin{equation}
    \text{AIC}= \chi^2_\text{min}+2k
\end{equation}
where $k$ denotes the number of free parameters, in order to account for the statistical penalty associated with model complexity. Differences in AIC are used to assess the relative preference between competing models. 

\subsubsection{Sampling CMB Anomaly Statistics}
\label{sec:method-2}
In this section, we describe the methodology adopted to assess the implications of IR cutoff models for large-scale CMB anomalies. Our primary focus is on the low quadrupole power, which constitutes one of the main motivations for introducing an IR cutoff in the primordial power spectrum. However, previous studies have shown that several large-scale CMB anomalies may be mutually correlated \cite{Muir:2018hjv, Jones:2023ncn}. Motivated by this, we also examine whether a reduction in the quadrupole amplitude can affect other prominent anomalous features, in particular the odd–even parity anomaly, or parity asymmetry \cite{Kim_2010}. Our objective is therefore to derive the posterior distributions of the statistics used to quantify these anomalies and to investigate their possible correlations within the framework of infrared cut-off features in the primordial power spectrum. For this anomaly-focused analysis, we employ derived large-scale statistics specifically designed to quantify known CMB anomalies. These statistics are typically non-Gaussian, low-dimensional summaries of the CMB maps or low-$\ell$ CMB spectra and are often analysed within a frequentist framework. Here, we adopt an approximate Bayesian approach, similar to that discussed in Refs.~\cite{Efstathiou_2010, Efstathiou:2003wr}, to assess these anomaly statistics.
\par We begin by focusing on the low CMB temperature quadrupole, $C_2$, and describe how we obtain the posterior distribution of the true theory quadrupole power, \UK{$C_2^{\text{th}}$}, given the observed value $C_2^{obs}$ from the data. Under the assumptions of statistical isotropy, Gaussian random initial conditions, full-sky coverage, and negligible instrumental noise, the observed angular power spectrum \UK{$C_\ell^{\text{obs}}$} provides a lossless summary of the CMB data. Adopting a uniform prior on $C_\ell^{\text{th}}$, the posterior distribution of the theoretical power spectrum is given by an inverse-gamma distribution,
\begin{equation} 
P\!\left(C_\ell^{\text{th}}\mid C_\ell^{\text{obs}}\right) \propto \left(\frac{C_\ell^{\text{obs}}}{C_\ell^{\text{th}}}\right)^{\frac{2\ell-1}{2}} \exp\!\left[-\frac{2\ell+1}{2} \left(\frac{C_\ell^{\text{obs}}}{C_\ell^{\text{th}}}\right)\right] \frac{1}{C_\ell^{\text{th}}} \, . 
\label{eq:inverse-gamma}
\end{equation}
Sampling from the above distribution, we obtain $P\!\left(C_2^{T}\mid C_2^{\text{obs}}\right)$ given the observed value of $C_2^{obs}$ from Planck 2018 data, yielding the posterior distribution of the true quadrupole amplitude conditioned on the data. To facilitate comparison with the theoretical expectations of different models, we also derive the posterior distributions of the model-predicted quadrupole amplitudes, $C_2^{(m)}$, where $m$ denotes a given cosmological model. These posteriors are obtained by treating $C_2^{(m)}$ as a derived parameter, evaluated using the {\tt CLASS} Boltzmann solver for each sample in the MCMC chains generated for the different models following the procedure described in Sec.~\ref{sec:method-1}. 
\par The next part is to look for other anomalies which we can, to some extent, relate to the angular power spectrum measurements of $C_\ell$'s. One notable large-scale feature observed in the CMB temperature anisotropies is the so-called odd--even parity anomaly or parity asymmetry \cite{Kim_2010}. It manifests as an anomalous excess of power in odd multipoles compared to even multipoles in the low-$\ell$ range ($2 \lesssim \ell \lesssim 30$). 
To quantify the odd--even asymmetry, one defines the parity asymmetry statistics $P$, given by the ratio of the mean power in even and odd multipoles \cite{Abdalla:2022yfr},
\begin{equation}
P \equiv \frac{P^{+}}{P^{-}},
\end{equation}
where
\begin{equation}
P^{\pm} = \sum_{\ell=2}^{\ell_{\max}} \frac{[1 \pm (-1)^{\ell}]\, \ell(\ell+1)\, C_\ell}{4\pi}.
\end{equation}
Here, $P^{+}$ and $P^{-}$ denote the total contributions from even and odd multipoles, respectively. 
Within the standard $\Lambda$CDM framework, statistical isotropy implies $P^{+} = P^{-}$, and hence $P = 1$. However, CMB observations from Planck 2018 yield $P<1$. While deviations from unity are often interpreted as potential violations of statistical isotropy, similar signatures can also arise from oscillatory features in the primordial power spectrum, such as those induced by IR cutoff models discussed in Sec.~\ref{sec:second}, which nevertheless preserve statistical isotropy. Following the same strategy adopted for the quadrupole analysis, we compare the predictions of the true theory for $P$ inferred from the data with those obtained from different inflationary power spectrum models.
\par In addition, we construct the joint posterior distributions of the anomaly estimators $C_2$ and $P$. Examining their joint behaviour across the sampled parameter space allows us to assess potential correlations between quadrupole suppression and parity asymmetry in the context of IR cutoff models.

\section{Results} 
\label{sec:fourth} 
In this section, we present the results of our analysis, organised into three subsections. We first examine infrared cut-off models of the primordial power spectrum using current observational data, with the aim of obtaining updated parameter constraints and assessing their viability relative to the standard $\Lambda$CDM model. We then investigate the connection of IR cut-off features on large angular scales with the reionisation optical depth ($\tau_\text{reio}$) and CMB–BAO discrepancy in light of recent DESI DR2 observations. Finally, we analyse the implications of these models for the low quadrupole anomaly and the parity asymmetry anomaly by deriving the joint posterior distributions of the corresponding anomaly statistics.

\subsection{Current Status of IR Cutoff Models}
\label{sec:results-1}
In this subsection, we present updated constraints on IR cutoff models using the recent datasets described in Sec.~\ref{sec: data}, following the methodology outlined in Sec.~\ref{sec:method-1}. In particular, we discuss results obtained for two data combinations: one using the CMB-only dataset, and another using the combined CMB$+$BAO$+$SN dataset, the details of which are summarised in Table~\ref{tab:datasets}. 
\par
\begin{figure}[h]
    \centering
     \includegraphics[scale=0.5]{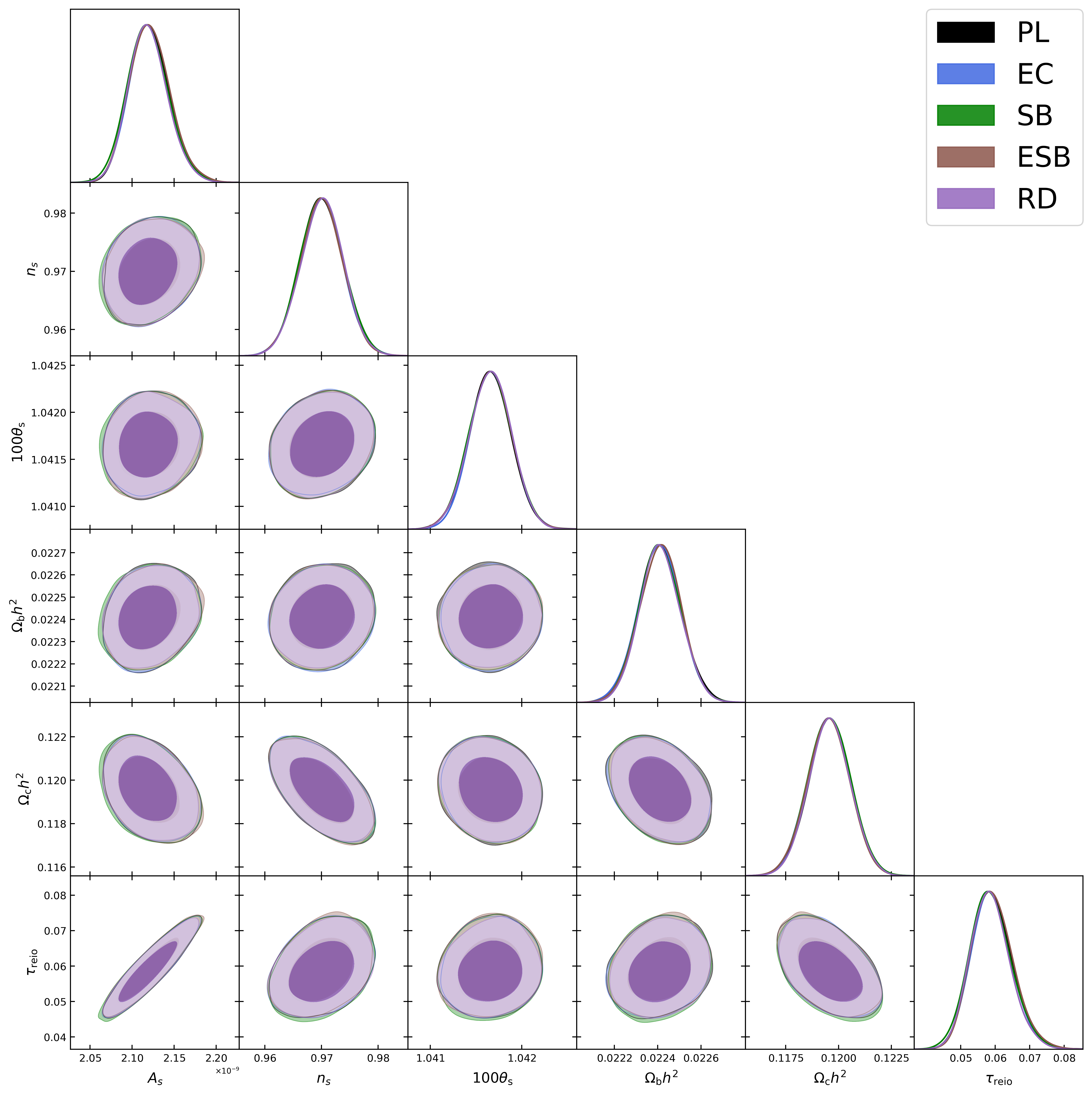}
     \caption{Constraints on IR-cutoff models using the full CMB dataset and the likelihood combinations listed in Table~\ref{tab:datasets}. The standard $\Lambda$CDM model with a power-law primordial spectrum is also shown for comparison.}
     \label{fig: cmb-constraints}
\end{figure}

\begin{figure}[h]
    \centering
     \includegraphics[scale=0.8]{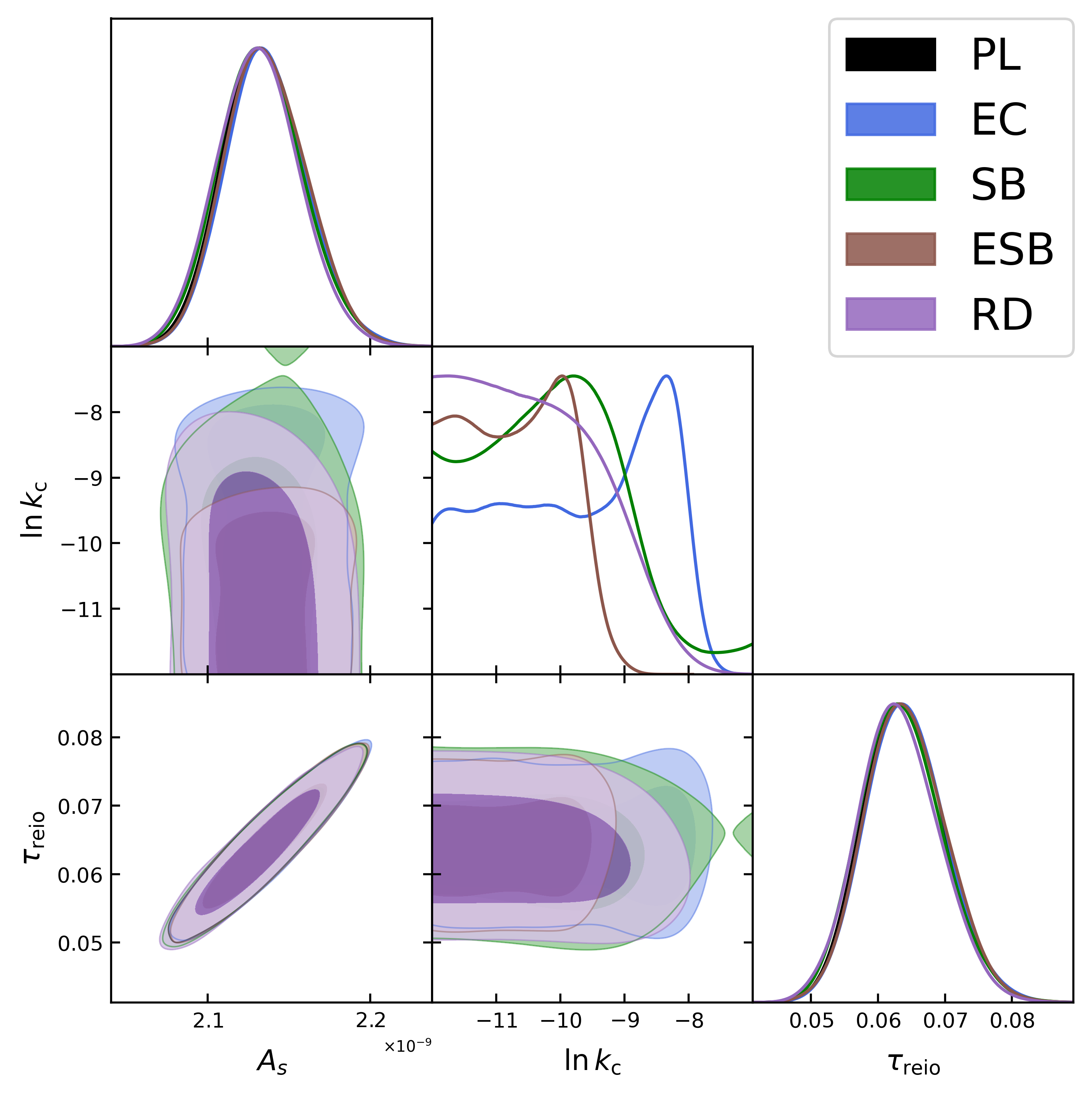}
     \caption{Constraints on IR cutoff parameter $k_c$ along with $A_s$ and $\tau_\text{reio}$ using the full CMB dataset and the likelihood combinations listed in Table~1. The standard $\Lambda$CDM model with a power-law primordial spectrum is also shown for comparison.} 
     \label{fig: ir-parameters-cmb}
\end{figure}

\begin{figure}[tbh]
    \centering
    \begin{minipage}[t]{0.48\textwidth}
        \centering
        \includegraphics[scale=0.50]{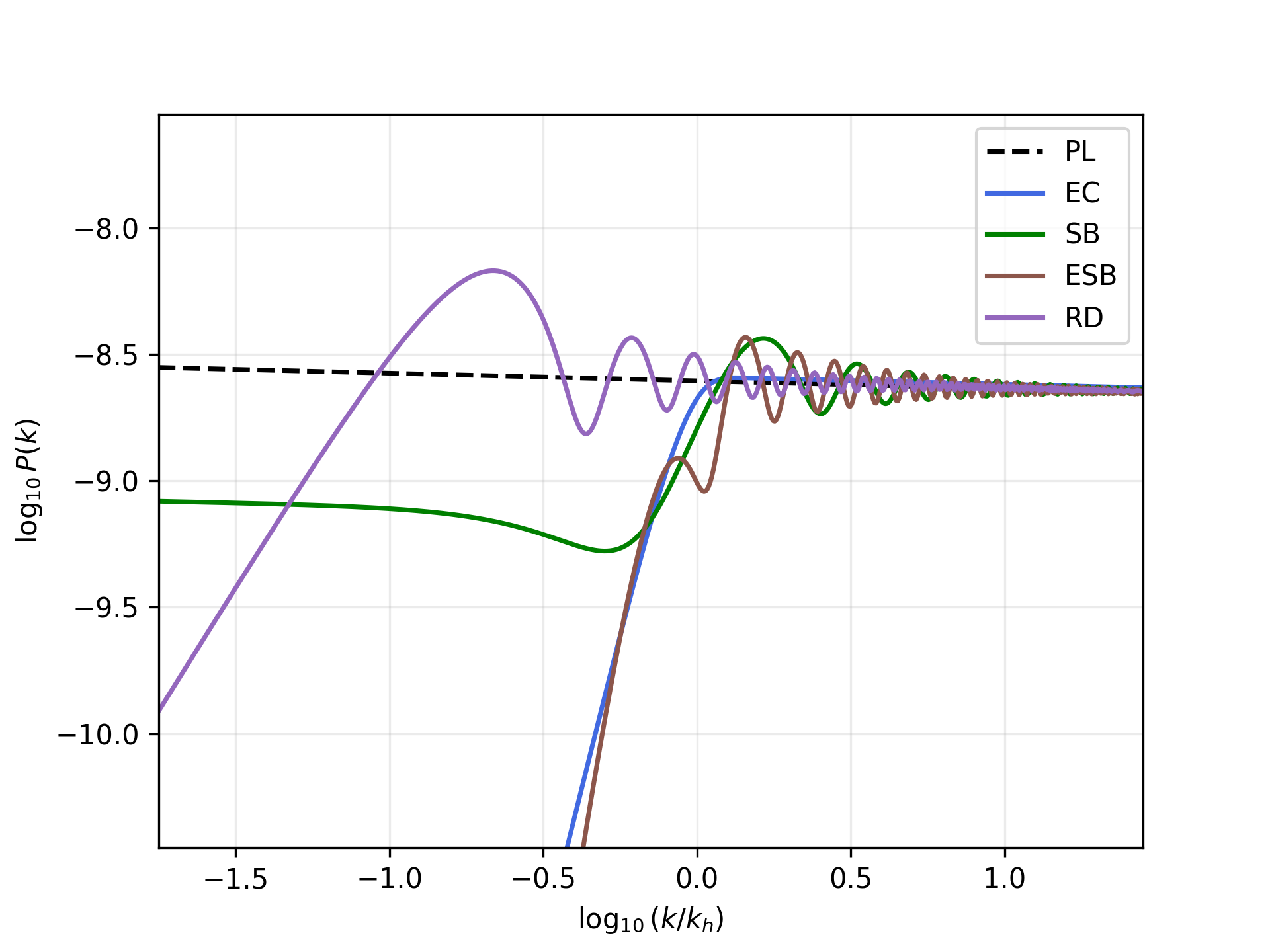}
        \subcaption{}
        \label{fig:IR}
    \end{minipage}
    \hfill
    \begin{minipage}[t]{0.48\textwidth}
        \centering
        \includegraphics[scale=0.50]{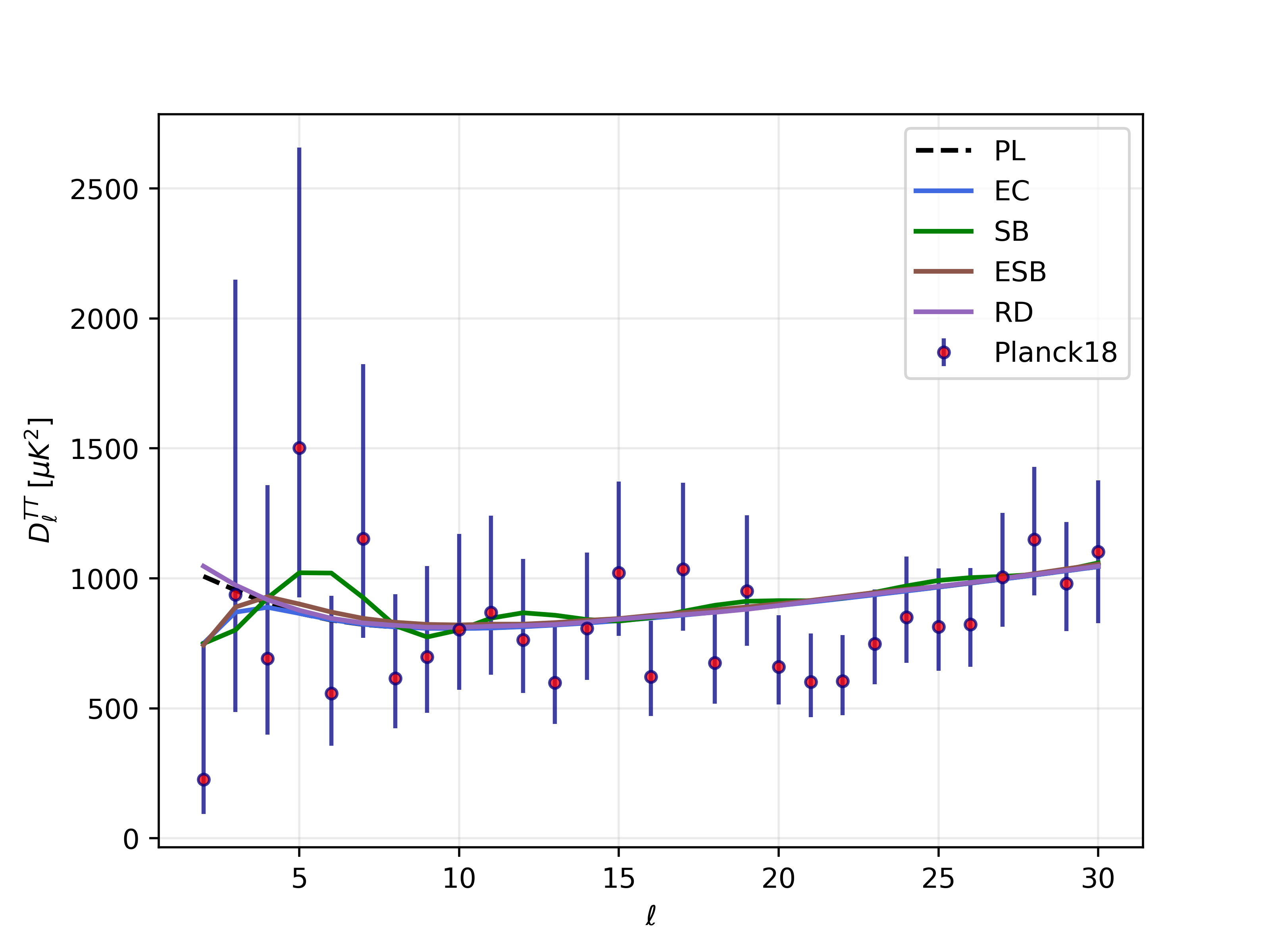}
        \subcaption{}
        \label{fig:EE}
    \end{minipage}
    \caption{Primordial power spectra for the IR cutoff models (left panel) and the corresponding imprints on the low-multipole CMB temperature anisotropy power spectrum (right panel), obtained using the best-fit parameter values listed in Table~\ref{tab: cmb-constraints} for the full CMB data. The reference power-law model is shown with a black-dashed line.}
    \label{fig:IR_bf}
\end{figure}

 Figure~\ref{fig: cmb-constraints} shows the marginalised posterior distributions of the six standard cosmological parameters for the IR cutoff models considered in this work, indicating the 68\% and 95\% credible regions. For comparison, the corresponding constraints obtained for the standard power-law primordial spectrum are also shown. We find that, for most models, the constraints on the six cosmological parameters remain largely unchanged and are consistent with those of the power-law case. 
Although IR cut-off features are expected to primarily affect the large-scale power and, consequently, parameters such as $A_s$, $n_s$, and $\tau_{\rm reio}$, we do not find any statistically significant deviations in these parameters when the models are confronted with the full CMB dataset. This can be attributed to the strong constraining power of the high-multipole measurements from the combined Planck, ACT, and SPT data, which dominate over the low-multipole information. These measurements, therefore, indirectly impose tight constraints on large-scale primordial features, such as the cut-off scale and the allowed level of power suppression at the largest angular scales. To illustrate this effect, Fig.~\ref{fig: ir-parameters-cmb} shows the constraints on the IR cutoff parameter $k_c$, which quantifies the cutoff scale in the different models, together with $A_s$ and $\tau_{\rm reio}$ obtained using the CMB-only dataset. In all cases, we obtain an upper bound on $k_c$, tightly constraining the position of the cutoff feature in the primordial power spectrum. In contrast, the cutoff scale remains largely unconstrained from below, reflecting the limited constraining power of data on the very largest scales. To further illustrate the impact of the constrained cutoff scale $k_c$, Fig.~\ref{fig:IR_bf} displays the primordial power spectrum and the corresponding CMB temperature anisotropy power spectrum at large angular scales for the best-fit parameter values of each model. As shown in the left panel of Fig.~\ref{fig:IR_bf}, the IR cutoff occurs at $k_c \leq k_h$ for all models considered. This, in turn, leads to a suppression of power in the CMB temperature anisotropy spectrum at \UK{very low multipoles, $\ell \sim 2$}, as illustrated in the right panel. The position and amplitude of the cutoff feature are therefore tightly constrained by the high-precision measurements of small-scale CMB power, which prevent IR cutoff models from producing arbitrarily large suppressions on observable large scales. In particular, for the RD model, the IR cut-off is pushed to extremely large scales, well beyond the present Hubble horizon. Consequently, the CMB angular power spectrum becomes effectively indistinguishable from the reference power-law case, exhibiting no appreciable suppression across the observable multipole range. This behaviour is also mirrored in the primordial power spectrum in the left panel of Fig. ~\ref{fig:IR_bf}, where the cut-off occurs at scales far outside those probed by current CMB observations. This result is consistent with the Grishchuk--Zel'dovich effect, which implies that the CMB anisotropy is primarily sensitive to power at low wavenumbers through the lowest multipoles, with the dominant contribution arising from the quadrupole \cite{Grishchuk:1978, Shafieloo:2003gf}. When the cut-off scale $k_c$ is extremely small, as in the RD model, the associated suppression lies at scales that project beyond the observable multipole window, leaving the low-$\ell$ CMB spectrum effectively unchanged.

Table~\ref{tab: cmb-constraints} summarises the best-fit parameter values for all the IR cutoff models, as well as for the reference power-law model, obtained from the MCMC analysis using the CMB-only dataset. We also report the goodness-of-fit estimates in terms of $\chi^2$, including the individual contributions from each CMB likelihood component listed in Table~\ref{tab:datasets}. Overall, all IR cutoff models provide fits to the data that are comparable to the power-law case, as indicated by the similar values of $\chi^2_{\rm total}$ across all models. We observe a modest improvement in the low-$\ell$ temperature anisotropy contribution to $\chi^2$ for the EC, SB, and ESB models, which can be attributed to the mild suppression features visible in the right panel of Fig.~\ref{fig:IR_bf}. However, these improvements do not translate into a statistically significant reduction in the total $\chi^2$. To further assess the relative performance of models with additional parameters, we employ the Akaike Information Criterion. \UK{The resulting $\Delta\mathrm{AIC}$ values, reported in Table~\ref{tab: cmb-constraints}, indicate that among all IR cutoff models, SB and RD are statistically comparable to the base power-law model, while EC and ESB are mildly disfavoured, primarily due to the penalty associated with their increased parameter complexity.} 
\par
\begin{figure}[h]
    \centering
     \includegraphics[scale=0.9]{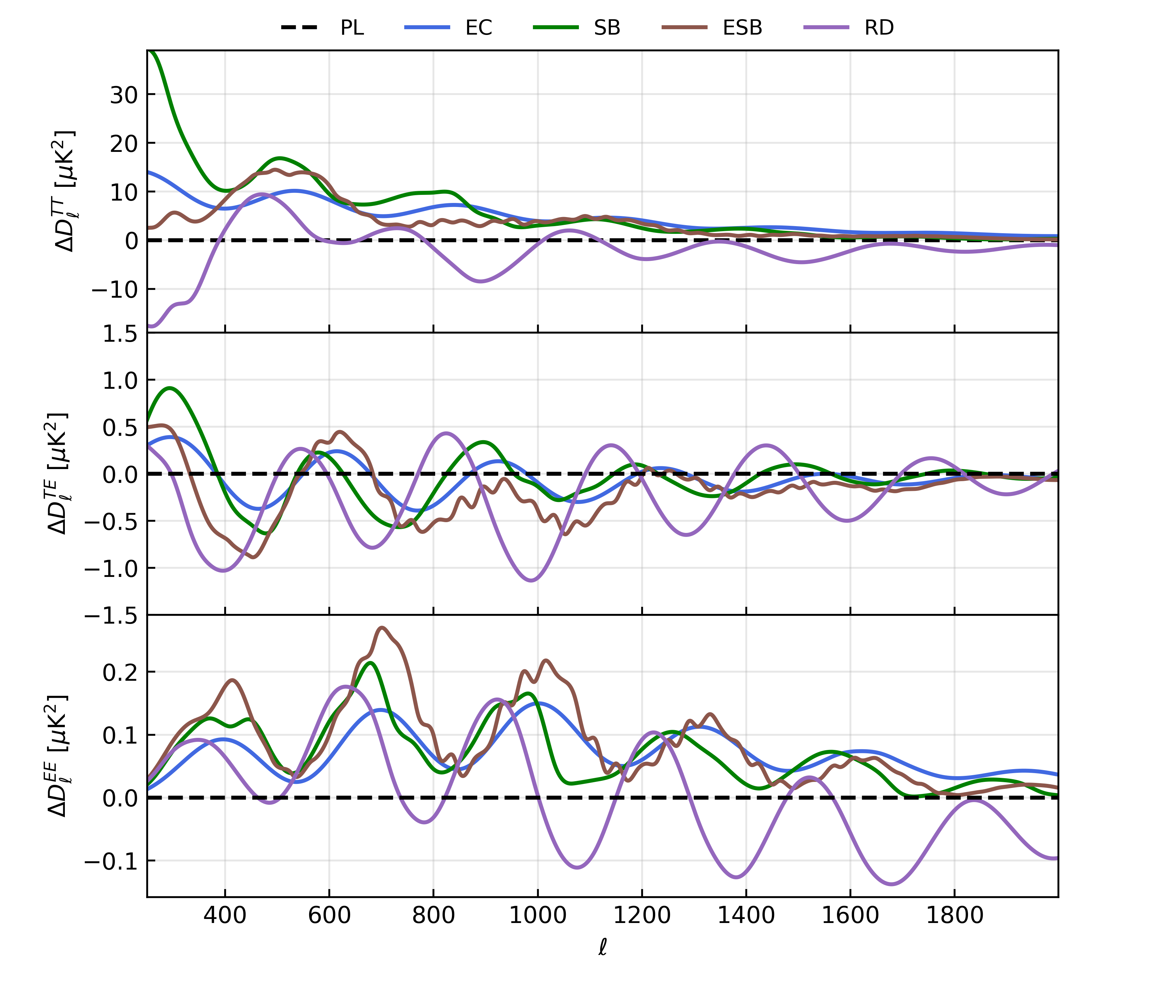}
     \caption{Best-fit residual CMB power spectra for temperature, polarisation, and their cross-correlation at large multipoles for the different IR cut-off models using parameter values from Table \ref{tab: cmb-constraints} obtained after MCMC analysis for the CMB-only data. The standard $\Lambda$CDM model corresponds to the zero line.}
     \label{fig: residual-CMB}
\end{figure}

 Another characteristic feature of some of the IR cut-off models is the presence of small-scale oscillations, which arise naturally in many phenomenological models of the primordial power spectrum that introduce large-scale power suppression, as illustrated in the left panel of Fig.~\ref{fig:IR_bf}. In our analysis, we include high-precision high-$\ell$ data from Planck, ACT, and SPT in order to constrain these oscillatory features, and we report the corresponding $\chi^2$ contributions in Table~\ref{tab: cmb-constraints}. We find, however, that the amplitude of the IR-induced oscillations on small scales is significantly smaller than the sensitivity of current observations in this multipole range, as shown in Fig.~\ref{fig: residual-CMB-withData} in Appendix~\ref{sec:appendix}, where we include the latest SPT data for comparison. To illustrate the impact of these features, we plot the residual CMB power spectra at high multipoles in Fig.~\ref{fig: residual-CMB}, computed using the best-fit parameter values of each model. From Fig.~\ref{fig: residual-CMB}, we find that, among all the models, the RD case provides an interesting exception: it exhibits a higher oscillation frequency and residuals that alternate between positive and negative values, in contrast to the other models for which the residuals remain predominantly positive over the multipole range $100\lesssim\ell\lesssim1200$. Although these features remain small, they induce mild distortions in the parameter posteriors, particularly for $\Omega_b h^2$, $\Omega_c h^2$, and $\theta_s$ (and hence $H_0$), which display slight asymmetries in their 68\% credible intervals, as reported in Table~\ref{tab:all-cmb-constraints}. As we discuss in the next section, this subtle asymmetry—originating from the high-frequency oscillatory features in the primordial and CMB power spectra of the RD model—leads to interesting implications that are not present in the other scenarios.

\begin{table}[tbh]
\centering
\vspace{0.2cm}
\label{tab:empty_table}
\renewcommand{\arraystretch}{1.4}
\setlength{\tabcolsep}{14pt}
\begin{tabular}{|l|ccccc|}
\hline\hline
Parameter & PL &EC & SB & ESB & RD\\
\hline
$10^9A_s$ &2.1079&2.1257&2.1013&2.1228&2.11\\
$n_s$    &0.9696&0.9714&0.9669&0.9726&0.9684\\
$\tau_{reio}$     &0.0564&0.0598&0.0525&0.0604&0.0579\\
$\Omega_b h^2$ &0.0224&0.0224&0.0224&0.0223&0.0224\\
$\Omega_c h^2$ &0.1198&0.1192&0.1197&0.1184&0.1199\\
$H_0\; \rm [km\;s^{-1}\;Mpc^{-1}]$ &67.4042&67.6197&67.4864&67.8463&67.3878\\
$10^4k_c\;\rm[Mpc^{-1}]$&-&2.1580&1.4056&0.4770&0.2561 \\
$\alpha$ &-&4.7596&-&4.8963&-\\
$R_*$ &-&-&0.5586&0.0422&-\\
\hline
$\chi^2_{\rm low-\ell\hspace{0.1cm}TT}$ &22.5389&21.2241&21.7607&20.9235&22.4575\\
$\chi^2_{\rm low-\ell\hspace{0.1cm}EE}$ &390.051&390.774&390.243&390.925& 390.229\\
$\chi^2_{\rm PlanckACT_\text{cut}}$ &221.198&221.0&218.637&219.208&221.368 \\
$\chi^2_{\rm ACT}$ &158.732&158.675&159.399&160.836&158.533 \\
$\chi^2_{\rm SPT}$ &160.967&161.612&161.681&161.662&160.802 \\
$\chi^2_{\rm PR4 \hspace{0.1cm}lensing}$ &8.7412&8.8353&8.5527&8.3760&8.7810 \\
$\chi^2_{\rm total} $ &962.228&962.12&960.274&961.93&962.171\\
\hline
$\Delta \chi^2$ & $0$ &  $-0.108$ & $-1.954$  & $-0.298$& $-0.057$\\
$\Delta \text{AIC}$ & $0$  & $+3.892$  & $+2.046$ & $+5.702$ & $+1.943$ \\
\hline\hline
\end{tabular}
\caption{Best-fit values of the parameters for the IR cut-off models obtained from the MCMC analysis using the full CMB-only data. The $\chi^2$ contributions from the individual likelihoods, along with the improvement in the total $\chi^2$ and the corresponding $\Delta$AIC with respect to the reference $\Lambda$CDM model with a power-law primordial power spectrum, are also reported.} 
\label{tab: cmb-constraints}
\end{table}

\begin{figure}[h]
    \centering
     \includegraphics[scale=0.5]{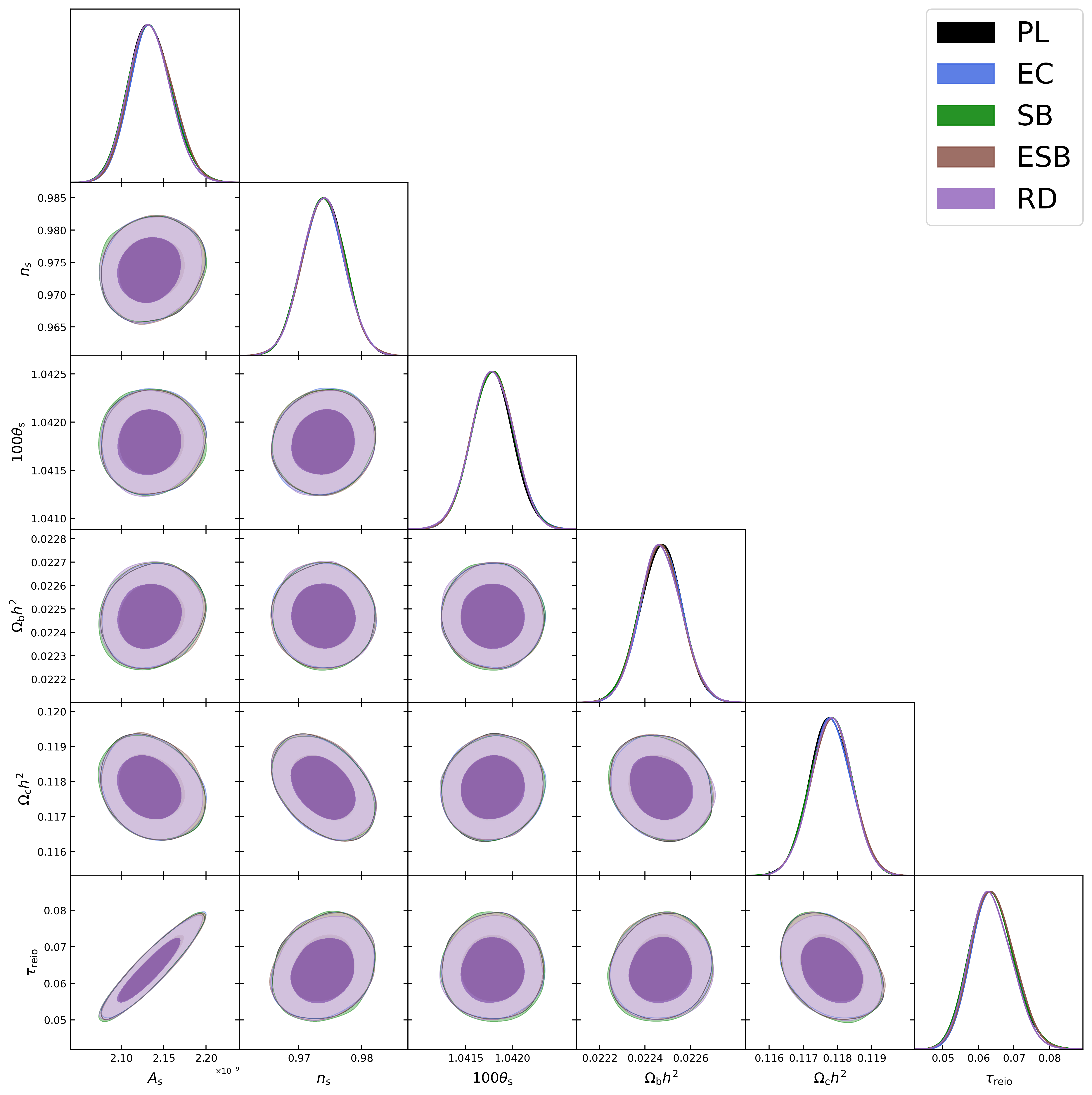}
     \caption{Constraints on IR-cutoff models using the full CMB+BAO+SNIa dataset and the likelihood combinations listed in Table~\ref{tab:datasets}. The standard $\Lambda$CDM model with a power-law primordial spectrum is also shown for comparison.}
     \label{fig: all-constraints}
\end{figure}
We next present the constraints on the model parameters obtained by combining BAO and supernova data with the CMB datasets, as described in Table~\ref{tab:datasets}. Figure~\ref{fig: all-constraints} shows the marginalised posterior distributions of the six standard cosmological parameters for all IR cutoff models, along with the reference power-law case. Consistent with the CMB-only analysis, we find that the constraints on the cosmological parameters remain largely unchanged relative to the power-law model.
\par For completeness, we also report the best-fit parameter values obtained from the joint CMB$+$BAO$+$SN analysis in \UK{Table~\ref{tab:all-bf}.} As in the CMB-only case, all models yield comparable values of $\chi^2_{\rm total}$, with no statistically significant improvement over the reference power-law model. While modest improvements in the low-$\ell$ temperature anisotropy contribution to $\chi^2$ are observed for the EC, SB, and ESB models, these do not translate into a meaningful reduction in the total $\chi^2$. \UK{Consequently, the resulting $\Delta$AIC values indicate that all IR cutoff models are mildly disfavoured, while RD remains statistically equivalent relative to the base power-law model, consistent with our earlier conclusions.}
\par In Appendix~\ref{sec:appendix}, we present the marginalised posterior distributions for the full parameter space of the IR cut-off models together with the six standard cosmological parameters, for both the CMB-only and the combined CMB+BAO+SN analyses. We also report the corresponding mean values and 68\% credible intervals for all parameters, along with selected derived cosmological parameters, for each IR cut-off model and for the reference power-law $\Lambda$CDM model.
\begin{table}[tbh]
\centering
\vspace{0.2cm}
\label{tab:empty_table}
\renewcommand{\arraystretch}{1.4}
\setlength{\tabcolsep}{14pt}
\begin{tabular}{|l|ccccc|}
\hline\hline
Parameter & PL &EC & SB & ESB & RD\\
\hline
$10^9A_s$ &2.1365&2.1428& 2.1302&2.1210&2.14\\
$n_s$    &0.9745&0.9733&0.9755&0.9735&0.9756\\
$\tau_{reio}$      &0.0641&0.0665&0.0604&0.0611&0.0644 \\
$\Omega_b h^2$ &0.0224&0.0224&0.0225&0.0224&0.0224 \\
$\Omega_c h^2$ &0.1175&0.1177&0.1176&0.1181&0.1176\\
$H_0\; \rm [km\;s^{-1}\;Mpc^{-1}]$ &68.2685&68.2728&68.2556&68.0568&68.3218\\
$10^4k_c \; \rm[Mpc^{-1}]$ &-&2.5556&1.3530&0.6410&0.1561 \\
$\alpha$ &-&5.0875&-&4.7338&- \\
$R_*$ &-&-&0.6047&0.0419&-\\
\hline
$\chi^2_{\rm low-\ell\hspace{0.1cm}TT}$ &21.8655&20.6799&20.1852&20.7269&21.8349\\
$\chi^2_{\rm low-\ell\hspace{0.1cm}EE}$ &391.563&392.959&391.96&391.587&391.688 \\
$\chi^2_{\rm PlanckACT_\text{cut}}$ &220.257&220.603&221.707&218.809&221.858 \\
$\chi^2_{\rm ACT}$ &160.639&160.45&159.924&160.549&160.076 \\
$\chi^2_{\rm SPT}$ &163.421&162.703&164.051&163.117&163.272 \\
$\chi^2_{\rm PR4 \hspace{0.1cm}lensing}$ &8.2776&8.2877&8.3012&8.3801&8.2970 \\
$\chi^2_{\rm DESI\hspace{0.1cm}DR2}$ &12.7231&12.5765&12.9391&15.1952&12.1958 \\
$\chi^2_{\rm PantheonPlus}$ &1405.66&1405.61&1405.6&1405.11&1405.73 \\
$\chi^2_{\rm total} $ &2384.41&2383.87&2384.67&2383.48&2384.95 \\
\hline
$\Delta \chi^2$ & 0 & $-0.54$  & $+0.26$  &$-0.93$ & $+0.54$\\
$\Delta \text{AIC}$ &0& $+3.46$ & $+4.26$  & $+5.07$ &$+2.54$  \\
\hline\hline
\end{tabular}
\caption{Best-fit values of the parameters for the IR cut-off models obtained from an MCMC analysis using the full CMB+BAO+SN data. The $\chi^2$ contributions from the individual likelihoods, along with the improvement in the total $\chi^2$ and the corresponding $\Delta$AIC with respect to the reference $\Lambda$CDM model with a power-law primordial power spectrum, are also reported.}
\label{tab:all-bf}
\end{table}
\subsection{IR Cutoff Models and CMB-BAO Discrepancy after DESI DR2}
\label{sec:results-2}
In this section, we investigate whether infrared cutoff features in the primordial power spectrum can be connected to the recently reported CMB–BAO discrepancy following the DESI DR2 results \cite{DESI:2025zgx}, and assess whether such inflationary scenarios offer any improvement over the standard power-law spectrum in this context. The expansion history inferred from DESI BAO measurements exhibits a mild tension with CMB-based predictions within the $\Lambda$CDM framework, most clearly visible in the $\Omega_m$–$H_0 r_d$ plane \cite{DESI:2025zgx}. This discrepancy has been interpreted as a possible indication of evolving dark energy rather than a cosmological constant \cite{DESI:2025zgx, DESI:2025fii}, or, in extended cosmological models, as a preference for values of the summed neutrino mass, $\Sigma m_\nu$, that are smaller than the lower bounds implied by neutrino oscillation experiments \cite{Elbers:2024sha, Elbers:2025vlz, Green:2024xbb, Craig:2024tky, Lynch:2025ine}. More recently, it has been argued that within the $\Lambda$CDM framework the optical depth to reionisation, $\tau_{\rm reio}$, may play a key role in alleviating these tensions if it is allowed to take values higher than those inferred from Planck CMB measurements \cite{Sailer:2025lxj, Jhaveri:2025neg, Planck:2018vyg}. This possibility is not unexpected and arises from the correlations of $\tau_{\rm reio}$ with other cosmological parameters, in particular the matter density $\Omega_m$ (in the present context), as discussed in detail in Refs.~\cite{Allali:2025yvp, Sailer:2025lxj}.
\par The reionisation optical depth, $\tau_{\rm reio}$, is predominantly constrained by large-scale CMB polarisation anisotropies sourced by Thomson scattering during the epoch of cosmic reionisation. This process generates additional polarisation power at very low multipoles ($\ell \lesssim 10$), producing the characteristic \emph{reionisation bump} in the $C_\ell^{EE}$ spectrum, whose amplitude scales approximately as $\tau_{\rm reio}^2$. Within the $\Lambda$CDM framework, the Planck 2018 analysis yielded a constraint of $\tau_{\rm reio} = 0.0544 \pm 0.0073$ \cite{Planck:2018vyg}. Subsequent SRoll2 reanalyses, benefiting from an improved control of large-scale polarisation systematics, reported a mildly higher value, $\tau_{\rm reio} = 0.0592 \pm 0.0062$, corresponding to an upward shift of approximately $0.5\sigma$ relative to the Planck 2018 legacy release \cite{Delouis:2019bub, Pagano:2019tci}.
Complementary information arises from the temperature anisotropy power spectrum, which constrains the combination $A_s e^{-2\tau_{\rm reio}}$ and thereby encodes the well-known degeneracy between the primordial scalar amplitude $A_s$ and the optical depth. The interplay between primordial features and reionisation signatures has recently been revisited in the context of IR cutoff models, motivated in part by the CMB--BAO inconsistency within $\Lambda$CDM discussed above \cite{Jhaveri:2025neg, Huang:2025xyf}. In these scenarios, a suppression of primordial power on the largest scales can partially counteract the enhancement of large-scale polarisation generated during reionisation, thereby biasing the inferred value of $\tau_{\rm reio}$ when confronted with CMB data within these cosmological frameworks. Relatedly, recent studies have also shown that relaxing the low-$\ell$ polarisation constraints—specifically by excluding large-scale $C_\ell^{EE}$ measurements—can also alleviate the CMB--BAO tension by relaxing the constraints on optical depth \cite{Sailer:2025lxj, Jhaveri:2025neg}.
\begin{figure}[tbh]
    \centering
    \begin{minipage}[t]{0.48\textwidth}
        \centering
        \includegraphics[scale=0.38]{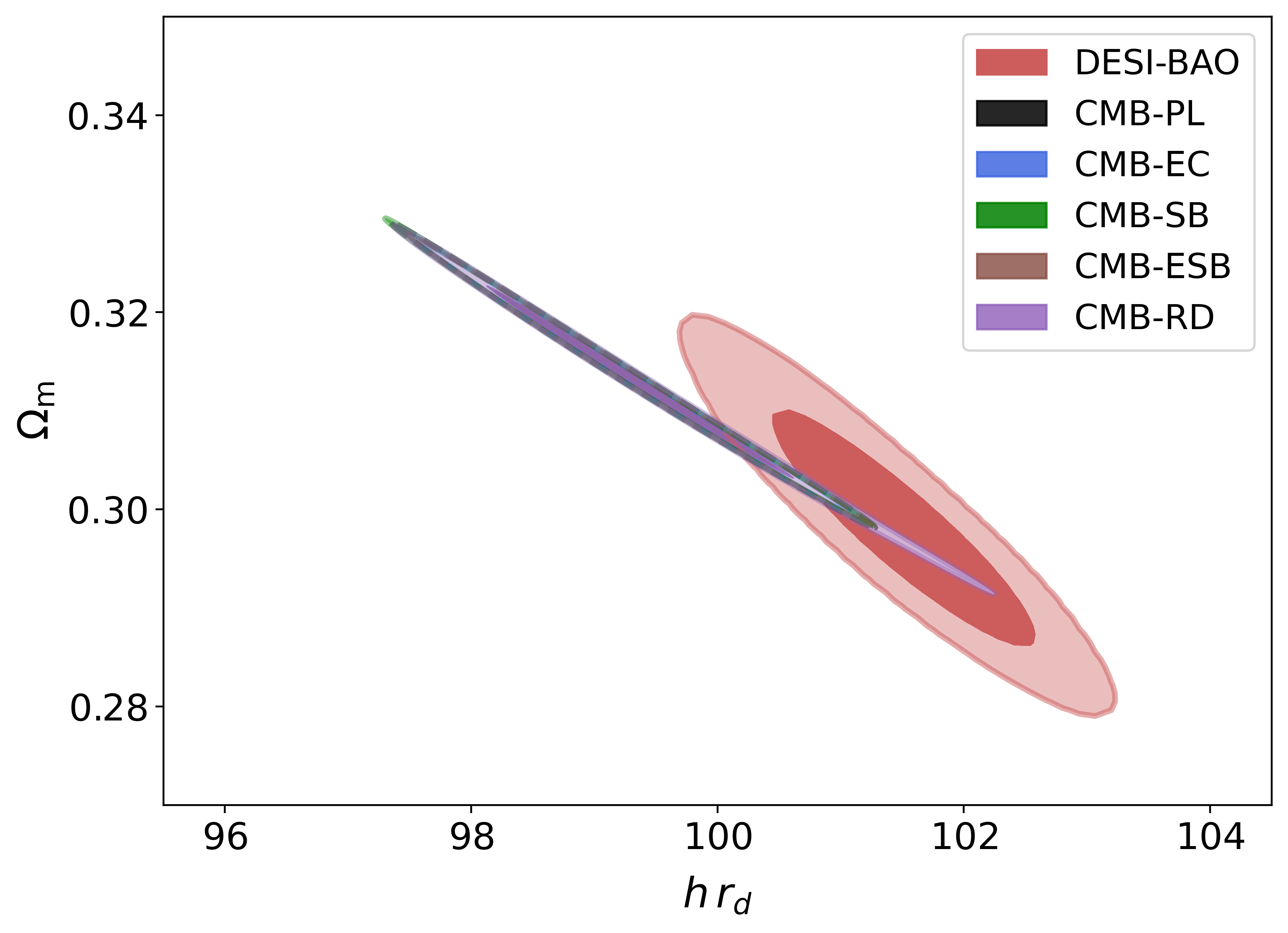}
        \subcaption{}
        \label{fig:with-EE}
    \end{minipage}
    \hfill
    \begin{minipage}[t]{0.48\textwidth}
        \centering
        \includegraphics[scale=0.38]{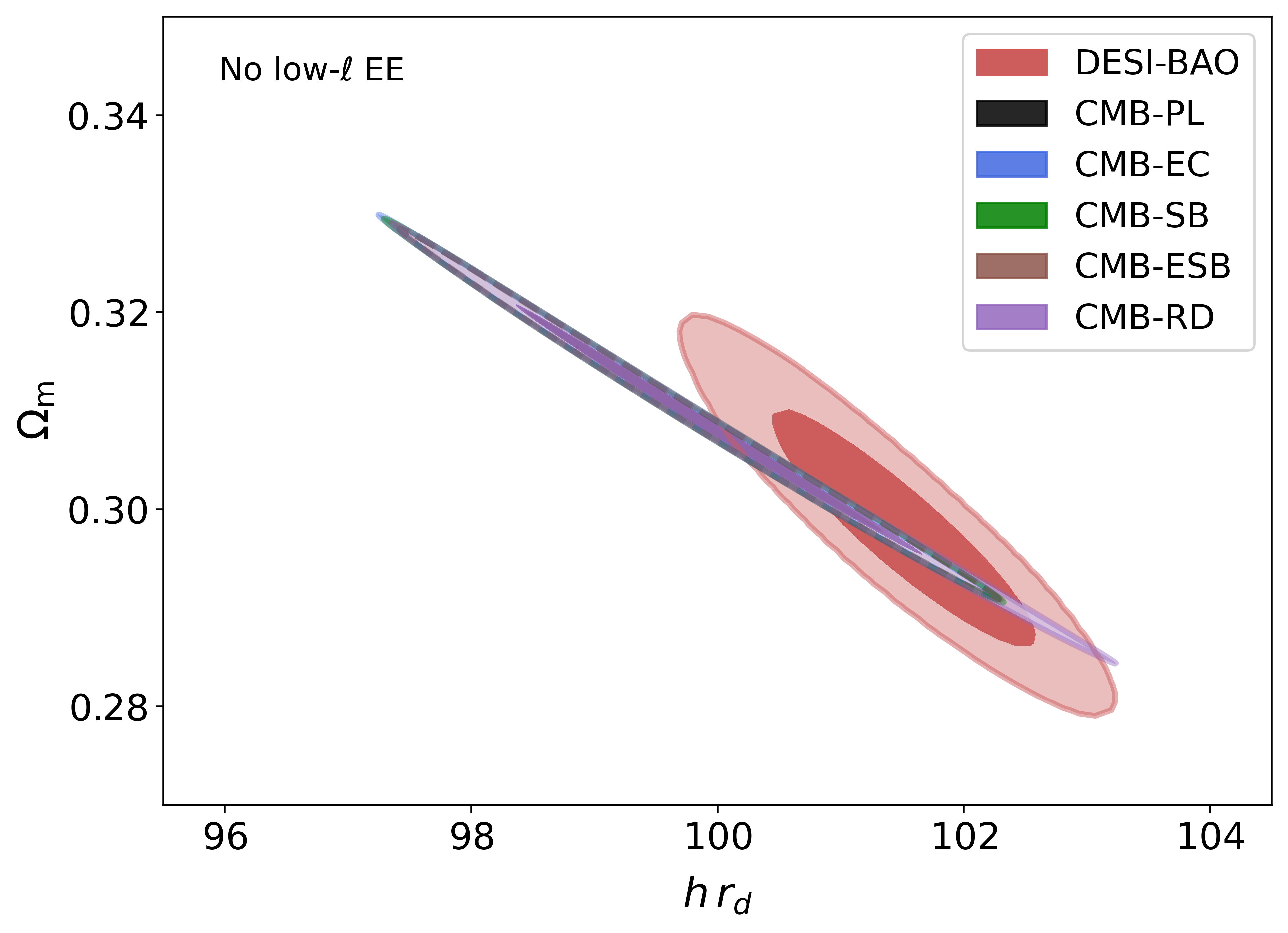}
        \subcaption{}
        \label{fig:no-EE}
    \end{minipage}
    \caption{Constraints in the $\Omega_m$--$h r_d$ plane for the IR cutoff models and the standard power-law $\Lambda$CDM scenario derived from CMB observations. The corresponding constraints from DESI BAO measurements are shown in red. The left panel displays results obtained using the full CMB data set, while the right panel shows the constraints when low-$\ell$ polarisation data are excluded. With the exception of the RD model, all IR cutoff scenarios yield constraints that closely track those of the reference power-law $\Lambda$CDM model, exhibiting a mild discrepancy with DESI BAO. Relaxing the polarisation constraints on $\tau_{\rm reio}$ leads to improved consistency with BAO in all cases, as shown in the right panel.
}
    \label{fig:CMB-BAO-1}
\end{figure}
\par In this section, we present a comprehensive analysis of the issues discussed above in light of recent DESI DR2 observations, focusing on IR cutoff models. For comparison, we also include the corresponding analysis within the standard power-law $\Lambda$CDM framework. Figure~\ref{fig:CMB-BAO-1} shows a comparison between the constraints derived from CMB data for the different models and those obtained from DESI BAO observations, projected onto the $\Omega_m$--$h r_d$ plane. The left panel of Fig.~\ref{fig:CMB-BAO-1} corresponds to constraints obtained using the full CMB data set, while the right panel shows the results when large-scale polarisation data are excluded to further relax the constraints on the reionisation optical depth. From the left panel of Fig.~\ref{fig:CMB-BAO-1}, it is evident that all IR cutoff scenarios yield constraints in the $\Omega_m$--$h r_d$ plane that are nearly identical to those obtained in the reference power-law $\Lambda$CDM case. With the exception of the RD model, all scenarios exhibit only a weak overlap with the DESI DR2 BAO constraints. We further note that this overlap is slightly improved in all models compared to the official DESI DR2 $\Lambda$CDM analysis \cite{DESI:2025zgx}. This difference arises primarily from our use of the SRoll2 likelihood for large-scale CMB polarisation, which shifts the inferred value of $\tau_{\rm reio}$ to slightly higher values and consequently modifies the inferred BAO consistency.
\par Among the IR cut-off models considered, the RD scenario exhibits a comparatively larger overlap with the BAO constraints, comparatively better than the standard power-law case. However, this apparent increase in consistency largely originates from the asymmetric error bars in the cosmological parameter constraints for the RD model, as reported in Table~\ref{tab:all-cmb-constraints}. These asymmetries propagate into the marginalised posterior in the $\Omega_m$–$h r_d$ plane, leading to an asymmetric contour that appears to overlap more strongly with the DESI DR2 constraints. 
This behaviour can be traced to the specific functional form of the primordial power spectrum in the RD model, which contains high-frequency oscillations extending over a wide range of scales. These oscillations imprint distinctive features in the CMB power spectrum, as shown in Fig.~\ref{fig: residual-CMB}, where the residual temperature spectrum alternates between positive and negative values over the relevant multipole range, in contrast to the other IR cut-off scenarios. This leads to small distortions in the posterior distribution of parameters $\Omega_bh^2$, $\Omega_ch^2$, $\theta_s$ and hence on derived quantities like $\Omega_M$ and $H_0$. This mechanism underlies the enhanced apparent consistency with BAO measurements in the RD case. 
\par Importantly, none of the IR cutoff models—aside from RD—outperform the standard power-law case, and they continue to exhibit a similar level of tension with BAO observations. This behaviour can be understood by noting that, in all IR cutoff models, the cutoff scale is tightly constrained by the CMB temperature power spectrum to very low multipoles, as also shown in the right panel of Figure \ref{fig:IR_bf}. Consequently, the associated suppression has a negligible impact on the large-scale polarisation anisotropy power spectrum, leaving insufficient freedom to accommodate a significantly higher value of $\tau_{\rm reio}$. This behaviour is also evident in Fig.~\ref{fig: cmb-constraints}, which shows that the constraints on $\tau_{\rm reio}$ for all IR cutoff models overlap almost perfectly with those obtained for the power-law case, indicating no significant shift in the inferred optical depth. Even in the case of the RD model, the improved consistency does not originate from the presence of a large-scale cut-off but due to a particular form of power spectrum as discussed above. Indeed, in this scenario, the cut-off scale $k_c$ is shifted to much smaller values, such that the model exhibits essentially no suppression at the largest observed multipoles, as shown in Fig.~\ref{fig:IR_bf}. 
\begin{figure}[tbh]
    \centering
     \includegraphics[scale=1.1]{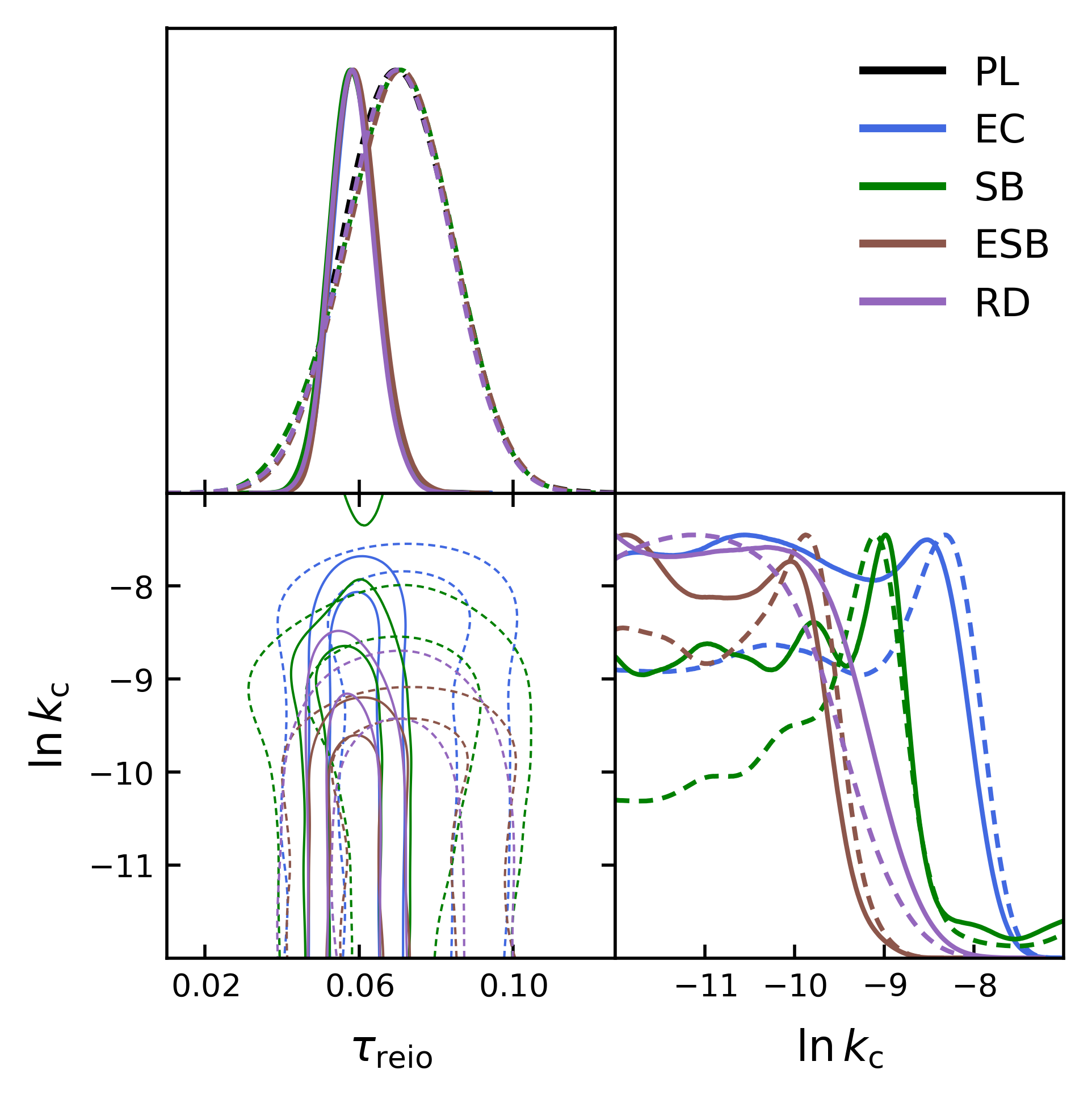}
     \caption{Joint posterior distributions in the $\tau_{\rm reio}$–$k_c$ plane for the various IR cutoff models. Solid contours correspond to constraints obtained using the full CMB data set, while dashed contours show the results when large-scale polarisation data (low-$\ell$ $C_\ell^{EE}$) are excluded. The constraint on $\tau_{\rm reio}$ from the reference power-law $\Lambda$CDM model is also shown for comparison (green). We find no significant shift in the cutoff scale $k_c$ upon removal of low-$\ell$ polarisation data, indicating that the cutoff parameter is primarily constrained by temperature anisotropies. In contrast, the shift in $\tau_{\rm reio}$ is driven by the relaxation of polarisation constraints and is common across all cosmological scenarios. Overall, this analysis shows that IR cutoff features do not significantly affect the inferred constraints on the reionisation optical depth.
  }
     \label{fig:CMB-BAO-2}
\end{figure}
\par In the right panel of Fig.~\ref{fig:CMB-BAO-1}, where the large-scale polarisation data are excluded (i.e.\ no low-$\ell$ $C_\ell^{EE}$), the consistency with BAO observations improves across all models, in a manner similar to the power-law $\Lambda$CDM case. This improvement is driven by the relaxation of constraints on the reionisation optical depth, $\tau_{\rm reio}$, and is common to all scenarios. The RD model continues to exhibit a comparatively larger overlap with the BAO constraints, consistent with the trend observed when the full CMB data set is used.
\par Figure~\ref{fig:CMB-BAO-2} presents the marginalised posterior distributions of the cutoff scale $k_c$ and $\tau_{\rm reio}$ corresponding to the two cases shown in Fig.~\ref{fig:CMB-BAO-1}. Solid lines denote constraints obtained using the full CMB data set, while dashed lines correspond to analyses excluding large-scale polarisation. As evident from the figure, removing the low-$\ell$ polarisation data shifts the mean value of $\tau_{\rm reio}$ towards higher values and significantly increases its uncertainty. This behaviour is common to all IR cutoff models and closely resembles that of the power-law case. Notably, this shift in $\tau_{\rm reio}$ is not driven by the presence of the IR cutoff feature itself as reflected in the posterior distributions of $k_c$, which show no significant change between the full-data and polarisation-excluded cases. This is expected, since constraints on $k_c$ are dominated by temperature anisotropy measurements, which are unaffected by the removal of large-scale polarisation data.
\par In conclusion, while relaxing the constraints on $\tau_{\rm reio}$ improves the consistency between CMB and DESI BAO observations, IR cutoff models do not offer any additional leverage beyond that of the standard power-law scenario. The stringent constraints imposed by CMB data on infrared cutoff features leave little scope for these models to induce a significant bias in the inferred value of the reionisation optical depth. This indicates that large-scale inflationary features are unlikely to reconcile the CMB--BAO discrepancy within the $\Lambda$CDM framework.

\subsection{IR Cutoff Models and CMB anomalies}
\label{sec:results-3}
In this section, we examine the implications of IR cutoff features for large-scale CMB anomalies, focusing on anomaly statistics within a Bayesian framework and following the methodology outlined in Sec.~\ref{sec:method-2}. Our analysis concentrates on two anomalies that are directly connected to the CMB power spectrum: the low quadrupole power and the parity asymmetry. The former also provides the primary motivation for IR cutoff models, and we therefore assess the predictions of these models for the corresponding anomaly statistics.
\begin{figure}[tbh]
    \centering
    \begin{minipage}[t]{0.48\textwidth}
        \centering
        \includegraphics[scale=0.7]{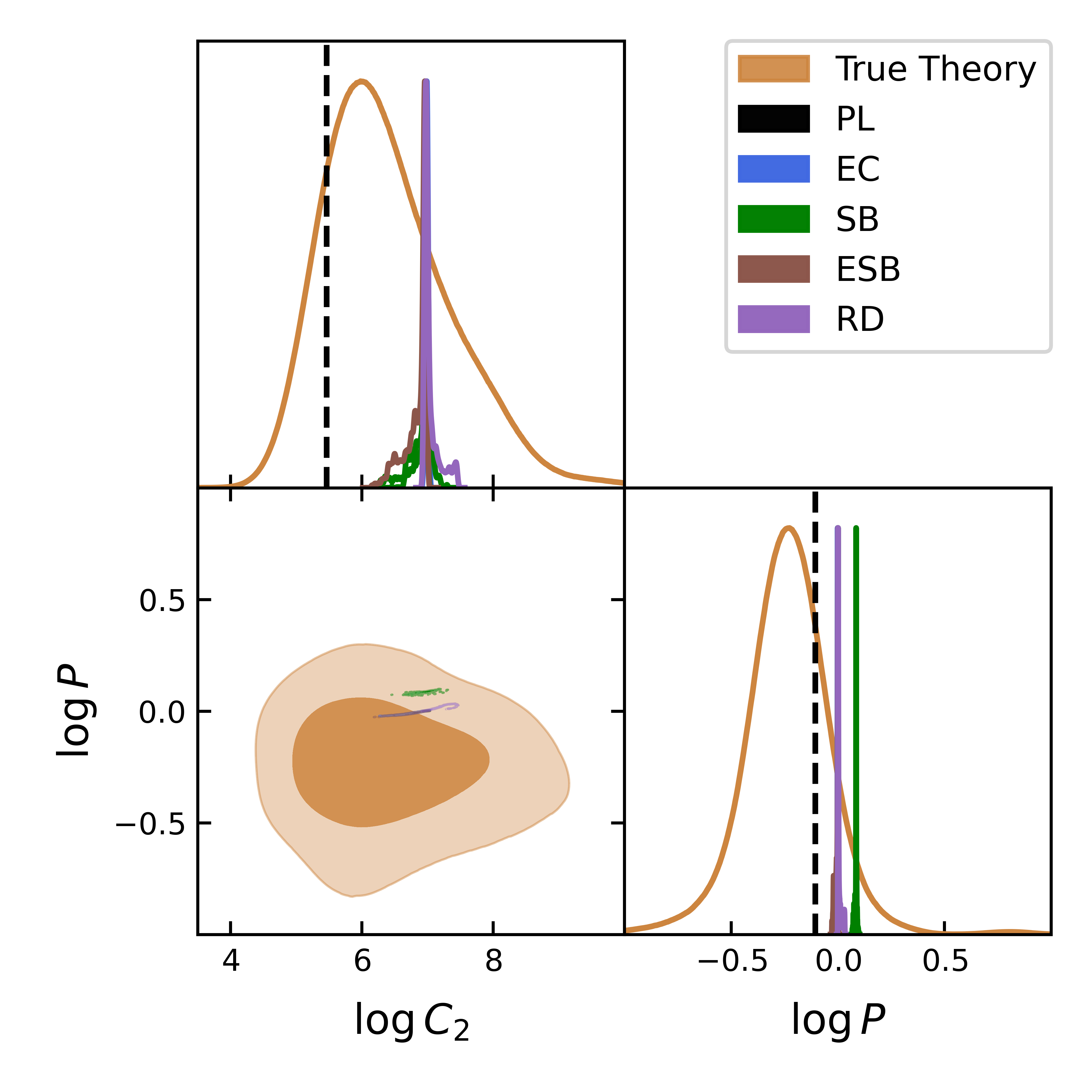}
        \subcaption{}
        \label{fig:with-observed}
    \end{minipage}
    \hfill
    \begin{minipage}[t]{0.48\textwidth}
        \centering
        \includegraphics[scale=0.7]{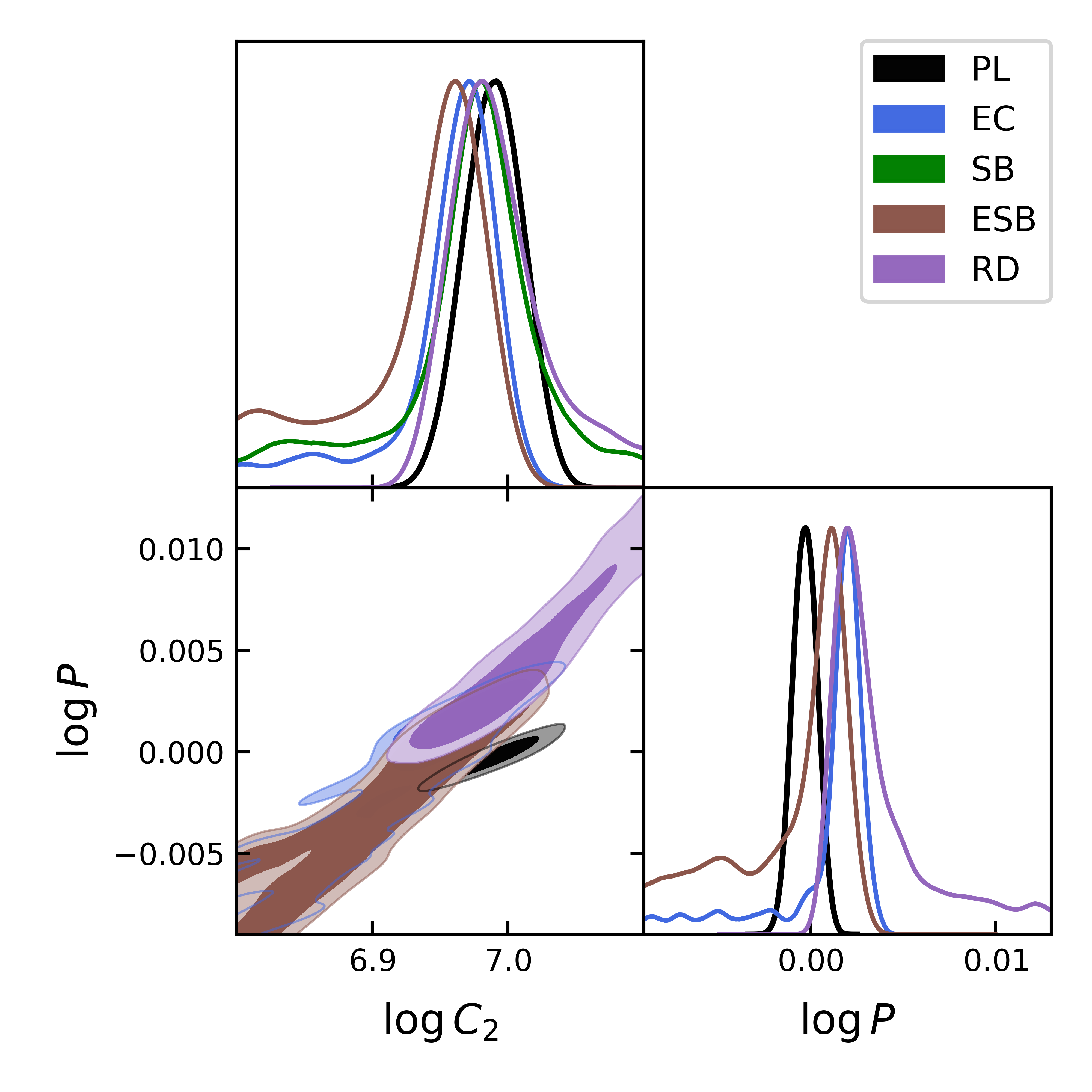}
        \subcaption{}
        \label{fig:with-models}
    \end{minipage}
    \caption{Posterior distributions of the anomaly statistics associated with the low quadrupole amplitude $C_2$ and the parity asymmetry statistic $P$. The left panel shows the joint posterior distribution for the ‘true theory’ conditioned on the observed sky under the assumption of statistical isotropy and Gaussian random initial conditions, together with the corresponding distributions inferred from the MCMC samples for the different cosmological models after parameter constraints from CMB data (shown in different colours). The right panel displays the posterior distributions for the cosmological models only, without the true-theory prediction for clarity. The black dashed lines in the one-dimensional distributions of $C_2$ and $P$ indicate their observed values from the Planck 2018 data.
 }
    \label{fig:CMB-anomalies}
\end{figure}
\par Figure~\ref{fig:CMB-anomalies} presents the joint posterior distributions of the anomaly statistics, namely the quadrupole amplitude $C_2$ and the parity asymmetry statistic $P$, thereby capturing their mutual correlations. In the left panel of Fig.~\ref{fig:CMB-anomalies}, the posterior corresponding to the `true theory' given the observations is shown in orange colour. This distribution is obtained by sampling the $C_\ell$ values from the inverse-gamma distribution described in Eq.~\ref{eq:inverse-gamma}, under the assumption of statistical isotropy and Gaussian random initial conditions, as discussed in Sec.~\ref{sec:method-2}. The coloured contours represent the posterior distributions of the anomaly statistics derived from the MCMC chains of the different cosmological models after their parameters have been constrained by CMB data. This framework provides a Bayesian perspective for assessing the predictions of different cosmological models by comparing their implied anomaly statistics with those expected from the true theory given the observed data, thereby enabling a qualitative assessment of their consistency with the observed large-scale anomalies.
\par As shown in the left panel of Fig.~\ref{fig:CMB-anomalies}, the posterior distribution corresponding to the `true theory' is broad and highly non-Gaussian, reflecting the large uncertainties induced by cosmic variance on these scales. This behaviour is particularly evident for the quadrupole amplitude \(C_2\), whose distribution follows the inverse-gamma form given in Eq.~\ref{eq:inverse-gamma}. The black dashed line marks the observed value of the CMB quadrupole from Planck 2018. By contrast, the predictions derived from the cosmological models are tightly constrained, since they are determined by the parameter uncertainties obtained from the MCMC analysis using the full CMB dataset. As a result, their posterior distributions appear as narrow vertical bands when overlaid on the `true theory' posterior. For this reason, in the right panel of Fig.~\ref{fig:CMB-anomalies} we present the posterior distributions of the anomaly statistics inferred for the different IR cut-off models, together with those of the reference power-law case, to facilitate a clearer comparison.
    \par The left panel of Fig.~\ref{fig:CMB-anomalies} shows that nearly all the models considered—both IR cutoff scenarios and the reference power-law case—are consistent with the predictions of the true theory given the observed data, with no statistically significant discrepancies. Notably, the mean values of the quadrupole amplitude distribution for the IR cutoff models are shifted slightly towards the observed value compared to the power-law case as can be seen in the right panel of Fig.~\ref{fig:CMB-anomalies}; nevertheless, all scenarios remain consistent with the true-theory posterior. 
    \par The Planck 2018 observations indicate a parity asymmetry statistic $P < 1$, as marked by the dashed vertical line in Fig.~\ref{fig:CMB-anomalies}, whereas the standard power-law $\Lambda$CDM case peaks close to unity. For the IR cutoff models, we find that the deviation of $P$ from unity is mildly enhanced but predominantly towards larger values, thereby shifting the predicted distributions away from the observed value and into the tail of the true-theory posterior. Among the models considered, the SB scenario exhibits the largest shift in $P$ relative to the observational value. This behaviour can be attributed to the oscillatory features at low multipoles inherent to this model, as illustrated in Fig.~\ref{fig:IR_bf}. We therefore find that, while all IR cut-off models predict a lower quadrupole amplitude $C_2$ than the power-law $\Lambda$CDM case, they simultaneously tend to favour $P>1$, moving away from the observed parity asymmetry. Nevertheless, these shifts remain small, and all models remain statistically consistent with the expectations of the true theory given the observations.

\par It is important to emphasise that the present analysis is carried out entirely within the framework of statistical isotropy and Gaussian random initial conditions. This distinction is particularly relevant because parity asymmetry, together with other large-angle anomalies such as the lack of large-angle correlations and the low northern variance, is often discussed in the context of possible violations of statistical isotropy \cite{Muir:2018hjv, Jones:2023ncn}. Since our study does not consider departures from statistical isotropy, we do not examine the implications for other anomaly statistics, such as $S_{1/2}$. Previous work has shown that these statistics are more directly linked to anisotropic models rather than to inflationary features that preserve isotropy \cite{Copi:2010na}.
\par In conclusion, within a Bayesian interpretation of anomaly statistics, the presence of large-scale anomalies, particularly the low quadrupole, does not constitute a significant discrepancy for the standard power-law description of the $\Lambda$CDM model. Although IR cutoff models can exhibit a marginally improved agreement, they are not statistically preferred once their increased model complexity is taken into account. Given the highly non-Gaussian nature of these anomaly statistics, a fully quantitative assessment of the relative compatibility of different cosmological frameworks with the true theory, as well as a robust evaluation of the statistical significance of these anomalies, is left for future work.

\section{Conclusion and Discussion}
\label{sec:five}
The primordial power spectrum is a crucial ingredient of the concordance $\Lambda$CDM framework specifying the initial conditions for cosmological perturbations that later evolve into the observed anisotropies in the CMB and the large-scale structure. Although it is not directly observable, the primordial spectrum can be inferred from measurements of the CMB power spectrum, which, however, probe a limited range of scales. As a result, modifications to the primordial power spectrum on the largest scales, such as IR cutoff features that lie beyond the most tightly constrained CMB modes, remain viable alternatives to the standard power-law description. These scenarios can lead to distinct phenomenology and have been motivated in part by large-scale CMB anomalies, most notably the observed suppression of power in the quadrupole.
\par In this work, we revisit a class of parametric models of the primordial power spectrum featuring an IR cutoff, in light of recent CMB observations. We present a comprehensive analysis of the phenomenology and implications of these scenarios in the context of several contemporary issues in cosmology. For clarity, the study is organised into three main parts, each addressing a distinct aspect of the impact of IR cutoff features. A common objective throughout is to assess the viability and relative performance of these models compared to the standard power-law case within a robust Bayesian framework.
\par In the first part of this work, we derive updated constraints on the cosmological parameters, together with the IR cutoff parameters in the different models, by fitting them to the latest CMB data from Planck, ACT, and SPT using a detailed MCMC analysis. We further incorporate complementary low-redshift information from DESI BAO and the Pantheon+ supernova data to obtain joint constraints on these scenarios. Finally, we compare the performance of the IR cutoff models with the standard power-law $\Lambda$CDM scenario using the Akaike Information Criterion, which assesses the goodness of fit while penalising models for increased complexity. We find that although certain IR cutoff models can yield a modest improvement in the chi-square of the large-scale temperature anisotropy data, reflecting a slight suppression of power around $\ell \sim 2$, the AIC analysis indicates that these models remain equally or less favoured than the standard power-law scenario once the penalty for additional parameters is taken into account. While the large-scale CMB data show a preference for suppressed power at low multipoles, as realised in IR cutoff scenarios, the strong constraining power of the higher-multipole CMB measurements tightly restricts the cutoff features. Consequently, the improvement at low multipoles is limited when the full CMB data set across all angular scales is considered in a robust statistical analysis. This result is expected and is consistent with previous analyses \cite{Sinha:2005mn}, while providing an updated picture in light of recent high-precision observations. \UK{While we have attempted to explore representative classes of IR cut-off models motivated by well-studied inflationary scenarios, it is important to emphasise that the primordial power spectrum remains, in principle, a functional degree of freedom. Consequently, forms of the PPS lying outside the classes considered here may still provide a significantly better fit to the data.}
\par In the second part of this work, we investigate possible connections between large-scale inflationary features and the reionisation optical depth, focusing in particular on whether the presence of such features can bias the inferred value of $\tau_{\rm reio}$ and allow for higher values compared to the standard power-law $\Lambda$CDM case. This question is motivated by recent studies suggesting that a larger optical depth could help reconcile the emerging tension between CMB and BAO measurements following the DESI observations within the $\Lambda$CDM framework. Our analysis shows that suppression features in the primordial power spectrum are tightly constrained by the CMB temperature anisotropy data, with the cutoff scale restricted to very low wavenumbers. As a result, only mild suppression around $\ell \sim 2$ is preferred. Consequently, the impact of these features on the large-scale polarisation power spectrum is negligible, leaving little room for degeneracies with $\tau_{\rm reio}$, which is otherwise strongly constrained by low-$\ell$ polarisation measurements. We therefore find that IR cutoff features do not provide any additional leverage to bias the inference of the reionisation optical depth relative to the standard $\Lambda$CDM scenario, owing to the limited preference for strong suppression on large scales in the full CMB data. 
\par Towards the end, we adopt a Bayesian approach to examine two large-scale CMB anomalies that are linked to the CMB power spectrum and hence to the IR cutoff features, namely the low quadrupole power and the odd--even parity asymmetry, quantified by the statistics $C_2$ and $P$, respectively as discussed in Sec. \ref{sec:method-2}. We derive the posterior distributions of these anomaly statistics for each cosmological model from the corresponding MCMC samples of the model parameters. 
We then construct the posterior distributions of the same statistics directly from the observed CMB data, under the assumptions of statistical isotropy and Gaussian random initial conditions, and compare them with the model-predicted posteriors. The uncertainty in the data-driven (`true-theory') posterior is dominated by cosmic variance, which plays a significant role on the largest angular scales on which these statistics are defined. In contrast, the posteriors derived from the cosmological models reflect uncertainties in the inferred cosmological parameters after fitting to the full CMB data set and are therefore considerably narrower. Our analysis shows that the posterior distributions corresponding to the true theory given the observations and those predicted by the various cosmological models largely overlap, indicating overall consistency between them. However, given the highly non-Gaussian nature of these anomaly statistics, we do not attempt a fully quantitative assessment of their statistical significance in this work, and instead leave such an analysis for future investigation.
\par IR cutoff models constitute an interesting class of inflationary scenarios that can arise from non-trivial physics operating at the largest cosmological scales. However, these scales are intrinsically difficult to probe with high precision because they are strongly limited by cosmic variance, which substantially weakens the constraining power of CMB observations at low multipoles compared to smaller angular scales. At present, measurements from Planck provide the most precise constraints on large-scale CMB anisotropies and are already close to the cosmic-variance limit. Although these data exhibit a mild preference for suppressed power on the largest scales, the statistical significance of this feature remains low. Consequently, current observations do not allow IR cutoff signatures to be distinguished from the standard power-law primordial spectrum with high confidence.

Looking ahead, future CMB experiments such as LiteBIRD \cite{2019BAAS51g.286L, LiteBIRD:2020khw} and \UK{CMB-Bh\=arat} \cite{cmb-bharat, Adak:2021lbu} are expected to deliver improved sensitivity on large angular scales, particularly through more precise measurements of large-scale polarisation \cite{Petretti:2024mjy}. While cosmic variance will continue to limit temperature anisotropies, enhanced control of systematics and improved polarisation measurements may provide complementary information and yield further insights into the viability of IR cutoff features in the primordial power spectrum.
\section*{Acknowledgement}
UU and YT thank Glenn Starkman and Matteo Braglia for helpful discussions. We also acknowledge Rajeev Kumar Jain for providing access to the PTG computing cluster at the Indian Institute of Science, which was used to carry out the computational analysis. UU and YT further thank Abhishek Tiwari for technical assistance with the installation of cosmological likelihoods and sampler code \texttt{cobaya} on the PTG cluster. \UK{TS acknowledges the J. C. Bose Grant of ANRF, India, under which YT was supported during this work.}
\appendix
\section{Constraints on the Full Parameter Space of IR cutoff models}
\label{sec:appendix}
\begin{figure}[tbh]
    \centering
     \includegraphics[scale=0.33]{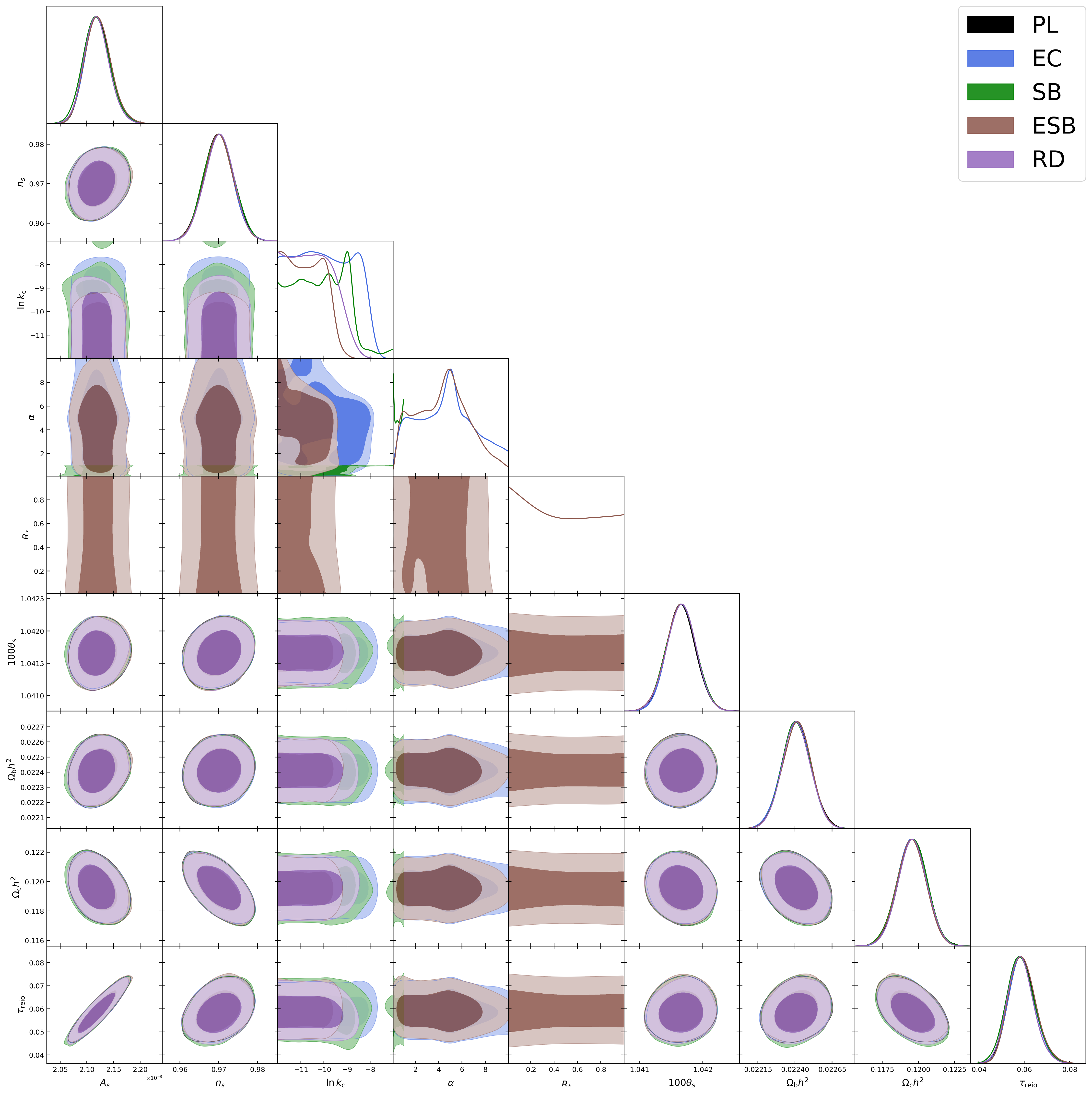}
     \caption{1D and 2D marginalised posterior distributions for the six cosmological parameters and all IR cut-off model parameters obtained from the CMB-only analysis.}
     \label{fig:all-params-CMB}
\end{figure}

In this section, we present the MCMC constraints on the full parameter space of the IR cut-off models together with the standard cosmological parameters, for both the CMB-only dataset and the combined CMB+BAO+SN dataset listed in Table~\ref{tab:datasets}. Figure~\ref{fig:all-params-CMB} shows the marginalised posterior distributions for the CMB-only case. The six standard cosmological parameters are tightly constrained, whereas the IR cut-off parameters remain largely weakly constrained. In all models, the cut-off scale $k_c$ exhibits an upper bound, confining the cut-off to very low multipoles. The remaining cut-off parameters, such as $\alpha$ and, in particular, $R_*$ in the EC, SB, and ESB models, show little or no meaningful constraint.
\par The corresponding 68\% credible intervals for all parameters are summarised in Table~\ref{tab:all-cmb-constraints}. For the SB and ESB models, no constraint is obtained on $R_*$, while $\alpha$ remains only weakly constrained. For the RD model, the posterior distributions appear approximately symmetric; however, the asymmetry in parameter constraints becomes evident when quoting the 68\% credible intervals. This behaviour originates from the specific functional form of the primordial power spectrum, which affects the CMB power spectrum in the multipole range where parameters such as $\Omega_c h^2$, $\Omega_b h^2$, and $\theta_s$ (or equivalently $H_0$) are most sensitive, as illustrated in Fig.~\ref{fig: residual-CMB}. As discussed in Sec.~\ref{sec:results-2}, these asymmetric constraints on the cosmological parameters, induced by features in the primordial power spectrum, lead to improved consistency between CMB and BAO measurements in the RD model, as shown in the left panel of Fig.~\ref{fig:CMB-BAO-1}.
\par
\begin{figure}[tbh]
    \centering
     \includegraphics[scale=0.33]{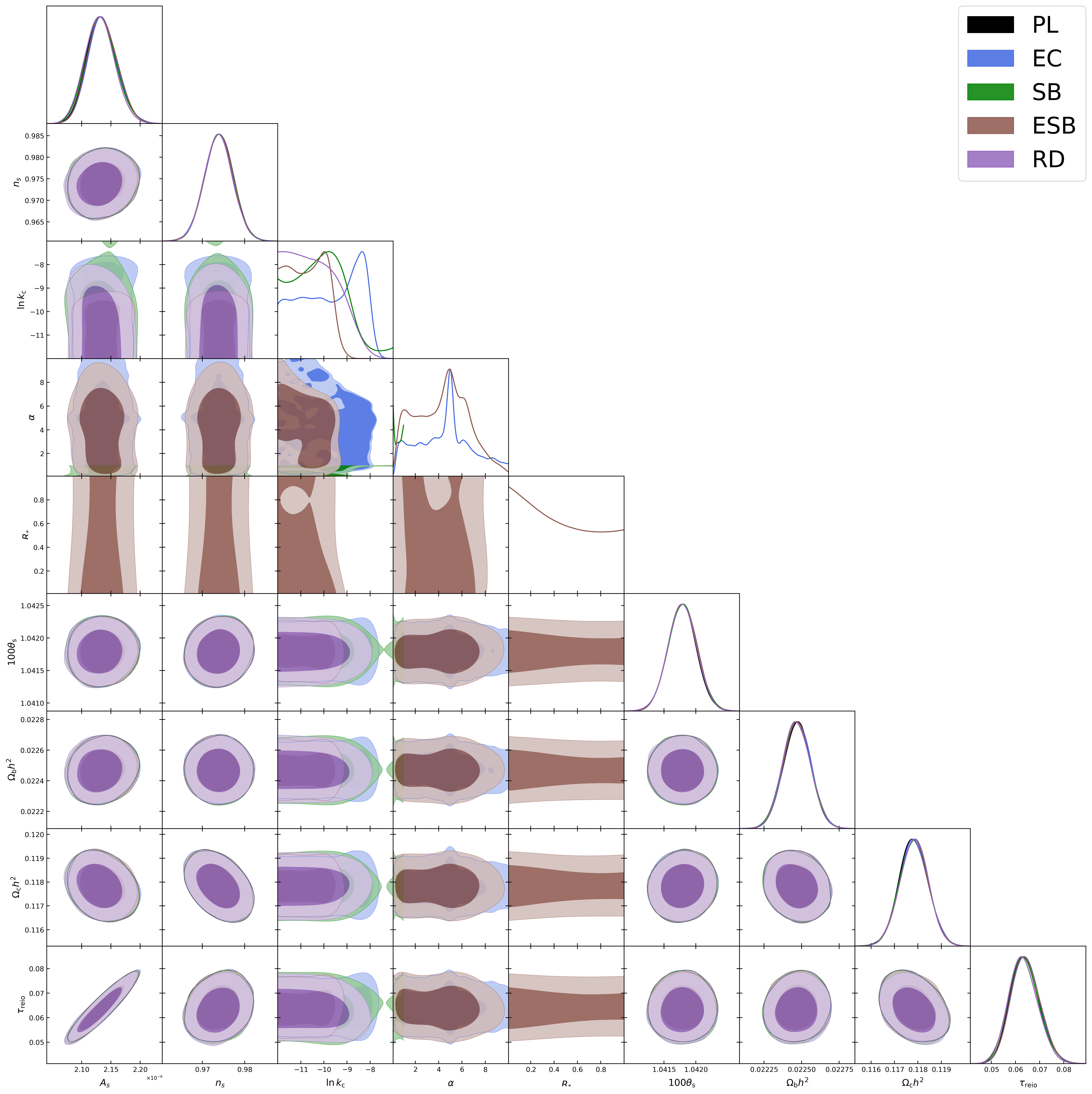}
     \caption{1D and 2D marginalised posterior distributions for the six cosmological parameters and all IR cut-off model parameters obtained from CMB + BAO + SNIa analysis.}
     \label{fig:all-params-all-data}
\end{figure}

 Figure~\ref{fig:all-params-all-data} shows the marginalised posterior distributions for the full data combination CMB+BAO+SN. The overall trends remain qualitatively similar to those obtained with the CMB-only dataset. In particular, the six standard cosmological parameters continue to be tightly constrained, indicating that the inclusion of BAO and SN data does not significantly alter their posterior distributions. Among the IR cut-off parameters, the cut-off scale $k_c$ again exhibits a clear upper bound, confining the cut-off to very low multipoles. The parameter $\alpha$ remains only weakly constrained, while $R_*$ continues to be unconstrained across the different models, reflecting the limited sensitivity of the current data to this parameter. The corresponding 68\% credible intervals for all parameters are summarised in Table~\ref{tab:all-all-constraints}.

\begin{table}[tbh]
\centering
\vspace{0.2cm}
\label{tab:empty_table}
\renewcommand{\arraystretch}{1.8}
\setlength{\tabcolsep}{0.95pt}
\footnotesize
\begin{tabular}{|l|ccccc|}
\hline\hline
Parameter & PL &EC & SB & ESB & RD\\
\hline
$10^9A_s$ &$2.121^{+0.021}_{-0.027}$&$2.121^{+0.022}_{-0.026}$&$2.114\pm 0.028$&$2.121^{+0.023}_{-0.026}$&$2.12^{+0.05}_{-0.08}$\\
$n_s$    &$0.9700\pm 0.0041$& $0.9699\pm 0.0040$ &$0.9698\pm 0.0043$&$0.9699\pm 0.0040$&$0.9723^{+0.0023}_{-0.0062}$\\
$\tau_\text{reio}$     &$0.0589^{+0.0053}_{-0.0063}$&$0.0590^{+0.0053}_{-0.0063}$& $0.0575\pm 0.0066$&$0.0594^{+0.0053}_{-0.0066}$&$0.0587^{+0.0055}_{-0.0068}$\\
$\Omega_b h^2$ &$0.02241^{+0.00010}_{-0.00010}$& $0.022408^{+ 0.000098}_{-0.000098}$&$0.02240^{ +0.00010}_{-0.00010}$&$0.022412^{+0.000099}_{-0.000099}$&$0.02241^{+0.00011}_{-0.00011}$\\
$\Omega_c h^2$ &$0.1196\pm 0.0011$& $0.1196\pm 0.0010$&$0.1197\pm 0.0011$&$0.1195\pm 0.001$&$0.1194^{+0.0012}_{-0.00099}$\\
$100\theta_s$ &$1.04166\pm 0.00023$& $4.8^{+2.2}_{-2.5}$&$1.04166\pm 0.00024$&$1.04165\pm 0.00024$&$1.04169^{+0.00022}_{-0.00026}$\\
\hline
$10^4k_c \;\rm [Mpc^{-1}]$ &-&$1.09^{+0.50}_{-1.10}$&$0.81^{+0.60}_{-0.79}$&$\,0.36^{+0.27}_{-0.31}\,$&$0.67^{+0.38}_{-0.63}$\\
$\alpha$ &-&$4.8^{+2.2}_{-2.5}$&-&$4.5^{+1.9}_{-2.4}$&-\\
$R_*$ &-&-&---&---&-\\
\hline
$H_0 \; \rm [km\;s^{-1}\;Mpc^{-1}]$ &$67.46\pm 0.43$&$67.49\pm 0.41$&$67.45\pm 0.44$&$67.49\pm 0.42$&$67.56^{+0.40}_{-0.51}$\\
$\Omega_m$ &$0.3136\pm 0.0062$&$0.3132\pm 0.0059$&$0.3138\pm 0.0063$&$0.3131\pm 0.0060$&$0.3122^{+0.0072}_{-0.0059}$\\
$\sigma_8$ &$0.8149^{+0.0044}_{-0.0051}$&$0.8146\pm 0.0050$&$0.8136\pm 0.0054$&$0.8146\pm 0.0050$&$0.8140\pm 0.0051$\\
\hline\hline
\end{tabular}
\caption{68\% credible intervals for the free cosmological parameters (first six rows), the IR cut-off model parameters (middle rows), and the derived parameters (last three rows) obtained from the MCMC analysis for the CMB-only dataset.}

\label{tab:all-cmb-constraints}
\end{table}

\begin{figure}[h]
    \centering
     \includegraphics[scale=0.8]{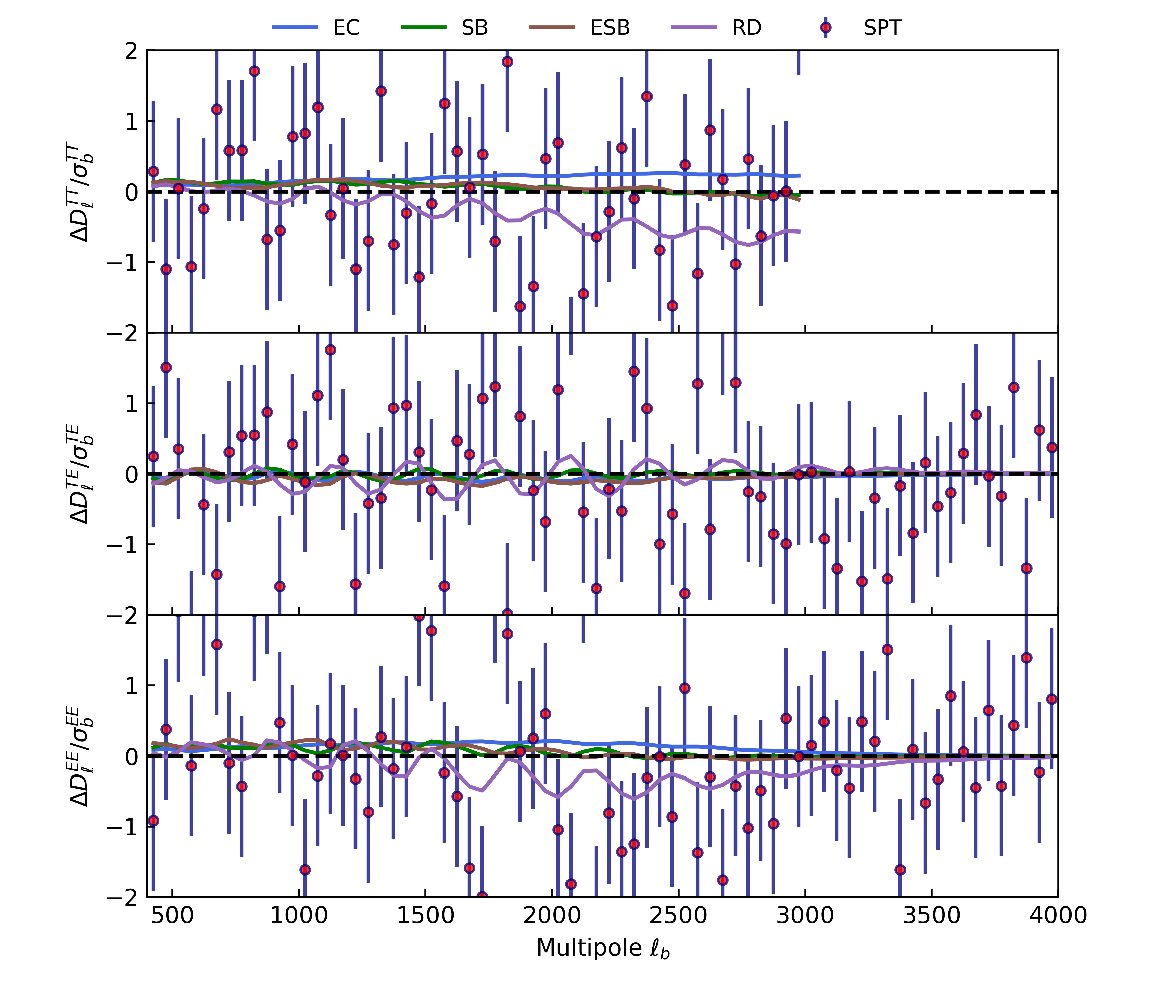}
     \caption{Best-fit residual CMB power spectra for temperature, polarisation, and their cross-correlation at large multipoles for the different IR cut-off models for the CMB-only data using parameter values from Table \ref{tab: cmb-constraints}. The standard $\Lambda$CDM model corresponds to the zero line, while the data points are from SPT observations \cite{SPT-3G:2025bzu}. For simplicity, ACT and Planck data are not included, since no features with significant amplitudes are found on these scales for the different IR cut-off models.}
     \label{fig: residual-CMB-withData}
\end{figure}

\begin{table}[tbh]
\centering
\vspace{0.2cm}
\label{tab:empty_table}
\renewcommand{\arraystretch}{1.8}
\setlength{\tabcolsep}{1pt}
\footnotesize
\begin{tabular}{|l|ccccc|}
\hline\hline
Parameter & PL & EC & SB & ESB & RD \\
\hline
$10^9A_s$        & $2.134\pm 0.027$ & $2.136^{+0.023}_{-0.027}$ & $2.131^{+0.025}_{-0.028}$ & $2.136\pm 0.027$  & $2.13^{+0.05}_{-0.07}$ \\
$n_s$            & $0.9738\pm 0.0039$ & $0.9739\pm 0.0037$ & $0.9738\pm 0.0035$  & $0.9736^{+0.0039}_{-0.0032}$  & $0.9758^{+0.0019}_{-0.0055}$ \\
$\tau_\text{reio}$ & $0.0640^{+0.0057}_{-0.0065}$  & $0.0640^{+0.0056}_{-0.0063}$ & $0.0630^{+0.0057}_{-0.0066}$ & $0.0642^{+0.0056}_{-0.0065}$ & $0.0627\pm 0.0068$ \\
$\Omega_b h^2$   & $0.022468^{+0.000098}_{-0.000098}$ & $0.022472^{+0.000093}_{-0.000093}$ & $0.022468^{+0.000095}_{-0.000095}$ & $0.022469^{+0.000096}_{-0.000096}$ & $0.02246^{+0.00010}_{-0.00010}$  \\
$\Omega_c h^2$   & $0.11785^{+0.00066}_{-0.00066}$ & $0.11785^{+0.00065}_{-0.00065}$  & $0.11785^{+0.00065}_{-0.00065}$ & $0.11787^{+0.00067}_{-0.00067}$ &  $0.11778^{+0.00070}_{-0.00061}$\\
$100\theta_s$    & $1.04178\pm 0.00023$ & $1.04179\pm 0.00023$ & $1.04179\pm 0.00023$ & $1.04179\pm 0.00023$ & $1.04181\pm 0.00025$ \\
\hline
$10^4k_c\;\rm [Mpc^{-1}]$            & - & $1.3\pm 1.2$ & $0.77^{+0.20}_{-0.74}$ & $0.39^{+0.15}_{-0.33}\,$ & $0.82\pm 0.83$ \\
$\alpha$         & - & $4.7^{+1.9}_{-2.4}$  &-  & $4.5^{+2.1}_{-3.9}$ & - \\
$R_*$            & - & - & --- & $< 0.539$ & - \\
\hline
$H_0 \; \rm [km\;s^{-1}\;Mpc^{-1}]$            & $68.19\pm 0.27$ & $68.20\pm 0.27$ & $68.19\pm 0.27$  & $68.19\pm 0.28$ & $68.22^{+0.25}_{-0.29}$ \\
$\Omega_m$       & $0.3032^{+0.0033}_{-0.0037}$ & $0.3031\pm 0.0036$ & $0.3032\pm 0.0037$ & $0.3033^{+0.0033}_{-0.0037}$ & $0.3028^{+0.0039}_{-0.0034}$  \\
$\sigma_8$       & $0.8123\pm 0.0053$ & $0.8127^{+0.0046}_{-0.0052}$ & $0.8117\pm 0.0051$  & $0.8126\pm 0.0055$ & $0.8117\pm 0.0056$ \\
\hline\hline
\end{tabular}

\caption{68\% credible intervals for the free cosmological parameters (first six rows), the IR cut-off model parameters (middle rows), and the derived parameters (last three rows) obtained from the MCMC analysis for the CMB+BAO+SN dataset.}
\label{tab:all-all-constraints}
\end{table}

\clearpage
\newpage

\bibliographystyle{JHEP}
\bibliography{references}

\end{document}